\documentclass[journal,12pt,onecolumn,draftclsnofoot]{IEEEtran}
\usepackage[numbers,sort&compress]{natbib}
\usepackage{amsmath,amssymb,amsfonts,amsthm,mathtools,bm}
\usepackage{graphicx}
\usepackage{float}
\usepackage{booktabs}
\usepackage{array}
\usepackage{enumitem}
\usepackage{needspace}
\usepackage{microtype}
\usepackage{xcolor}
\usepackage[colorlinks=true,linkcolor=blue,citecolor=blue,urlcolor=blue]{hyperref}

\newtheorem{theorem}{Theorem}
\newtheorem{proposition}{Proposition}

\newtheorem{corollary}{Corollary}

\newtheorem{remark}{Remark}

\newcommand{\X}{\mathcal X}
\newcommand{\Y}{\mathcal Y}
\newcommand{\Z}{\mathbb Z}
\newcommand{\R}{\mathbb R}
\newcommand{\E}{\mathbb E}

\newcommand{\pos}[1]{\left[#1\right]_+}
\newcommand{\1}{\mathbf 1}
\newcommand{\RR}{\mathrm{RR}}
\newcommand{\diag}{\operatorname{diag}}
\newcommand{\tr}{\operatorname{tr}}

\title{Support-Constrained Local Randomizers for Robust Private Distribution Learning}

\author{
\IEEEauthorblockN{Amirreza Zamani, Parastoo Sadeghi $^\dagger$, Mikael Skoglund \vspace*{0.5em}
		\IEEEauthorblockA{\\
		    \normalsize Department of Information Science and Engineering, KTH Royal Institute of Technology \\
			$^\dagger$School of Engineering and Technology, UNSW, Canberra, Australia\\
			Email: \protect amizam@kth.se, p.sadeghi@unsw.edu.au, skoglund@kth.se }}
            }
\begin{document}
\maketitle

\begin{abstract}
Local differential privacy protects users before data collection, but full-support randomizers can be wasteful on ordered, metric, or semantic alphabets. We study support-constrained local randomizers for robust private discrete learning. Restricting reports to input neighborhoods reduces reporting distortion but creates support mismatch. Under range-restricted approximate local differential privacy, a relational guarantee weaker than ordinary all-pairs local differential privacy, we decompose the exact hockey-stick defect into support leakage and overlap excess loss. For line-supported sparse randomized response, we derive the exact defect, solve the joint privacy-distortion problem in closed form at each radius, and prove a privacy-imposed support lower bound. For distribution estimation, we obtain an exact finite-sample risk identity and show that the least-favorable law over any closed convex uncertainty set is a Euclidean projection. On cyclic alphabets, the uniform law is uniquely least favorable, nominal and minimax mechanism rankings coincide, and the minimax risk obeys spectral upper and lower bounds. This diagnoses why prior shift cannot alter the selected mechanism in a homogeneous cyclic experiment. We therefore introduce a boundary-normalized nonhomogeneous path family in which robust calibration changes the mechanism and reduces worst-case risk. We also derive exact cyclic privacy, robust privacy-range calibration, and precise conditions for safe context- or history-dependent channel selection. Sparse Laplace, Gaussian, staircase, and score-based mechanisms remain extensions of the same design principle. All numerical results use exact privacy and risk calculations with matched-guarantee baselines.
\end{abstract}


\section{Introduction}

Information-theoretic methods provide a natural language for robust and adaptive learning because they quantify how data-processing systems acquire, distort, preserve, and protect information. In privacy-preserving learning, the local model is especially stringent: each user must randomize their sample before collection. Classical randomized response and its $k$-ary variants are full-support local randomizers, assigning positive probability to every output symbol. On ordered, metric, or semantic alphabets this can be statistically and operationally wasteful: if the private input is a count, a location bin, a rating, or a semantic category, only a local neighborhood of the input may be useful for the downstream task.

This paper studies support-constrained local randomization as a mechanism-design problem for robust private discrete learning. The mechanism is allowed to output only from an admissible neighborhood $S(x)$ around the private input $x$. Such a constraint can reduce reporting distortion, but it changes the privacy accounting. It reduces communication only when the decoder has an appropriate public anchor or other side information that permits a local offset representation; support restriction alone does not imply fewer transmitted bits. If an output belongs to $S(x)$ but not to $S(x')$, then observing that output rules out $x'$. Pure all-pairs local differential privacy (LDP) therefore forces common support. Approximate LDP permits input-dependent supports, but charges the mismatch to the $\delta$ budget through support leakage.
The central design question is not merely whether a sparse support is private, but how the support should be chosen when the learning environment may change. A deployed private learner may face changing public domains, changing privacy ranges, changing uncertainty about the population distribution, and changing metric structure. In such settings, the support radius is not an implementation detail; it is a design variable that balances reporting distortion, privacy leakage, and statistical risk. Our adaptation claim is deliberately narrow: a channel may be selected from genuinely public context, or from prior privatized history that is independent of a fresh user's current record. Tuning on the current unprotected private value is not free, and repeated participation requires privacy composition. We develop this viewpoint for sparse randomized response and then extend the same optimization language to Laplace, Gaussian, staircase, and score-based sparse mechanisms.\vspace{-3mm}
\subsection{Positioning and Novelty}
This paper is positioned at the intersection of local privacy, support-constrained mechanism design, and robust discrete learning with safe context-dependent channel selection. Existing sparse-LDP work has largely treated sparsity either as a property of the unknown population, as in sparse distribution estimation, or as a privacy--sparsity accounting question for particular kernel families. We instead treat the output support itself as a design variable: the mechanism chooses how large a local neighborhood to reveal, how much probability to allocate to the truthful symbol, and how these choices should change under uncertainty about the learning environment.
The novelty is therefore not only the support-leakage identity used for privacy accounting. That identity is the starting point. The main contribution is to turn support leakage into an optimization principle for private learning: for sparse randomized response we jointly optimize the support radius and reporting probabilities, derive exact inverse-estimation risk, identify least-favorable distributions analytically, and then extend the design to distributionally robust calibration, privacy-range mismatch, and safe public-context adaptation. This gives a robust-learning interpretation that is distinct from a broader sparse-kernel taxonomy: the support radius is a statistical design parameter rather than only a sparsity descriptor.
Compared with earlier sparse Laplace/Gaussian mechanisms and sparse point-wise privacy leakage formulations \citep{zamani2026sparseLDP,zamani2026sparsepointwise}, the present paper focuses on a different problem. It does not aim to classify all sparse local channels. Instead, it studies a concrete support-constrained randomizer deeply enough to obtain closed-form design rules, exact finite-sample distribution-estimation risk, spectral structure on cyclic alphabets, distributionally robust calibration, and safe context-selection rules. A concise discussion explains how the same design principle transfers to Laplace, Gaussian, staircase, and score-based channels, while the main theory remains centered on optimized private learning.

\subsection{Related Work and Connections to Sparse Mechanisms}

Differential privacy was introduced in \citep{dwork2006calibrating}, and the local model and its statistical consequences are discussed in \citep{dwork2014book,duchi2014local,duchi2018minimax}. Randomized response originates with Warner \citep{warner1965randomized} and remains fundamental in local distribution estimation and systems such as RAPPOR \citep{rappor2014,kairouz2016discrete,pastore2021randomized}. Sparse distribution estimation under LDP shows that intrinsic support size can govern sample complexity \citep{acharya2021sparse}. Utility- and context-aware LDP mechanisms design channels for improved inference \citep{murakami2019utility,acharya2020context}, while staircase and score-based mechanisms provide general templates for privacy--utility optimization \citep{mcsherry2007mechanism,geng2014optimal,kairouz2014extremal}.

Robustness and adaptation under LDP have also been studied for contamination, evolving data, and online distribution learning \citep{chhor2023robust,li2023robustness,joseph2018evolving,sima2024online}. Those works address statistical robustness or temporal learning protocols. Here robustness concerns uncertainty in the deployment distribution and in the protected input range, with the channel itself optimized through exact finite-sample inverse risk. Our context-dependent result is correspondingly more limited than a general online-learning claim: it specifies when selecting among pre-certified local channels preserves the advertised privacy guarantee.

Information-theoretic formulations of privacy through divergences and hypothesis tests are also closely related. Asoodeh et al. develop lossless conversions between approximate DP, Renyi DP, and hypothesis-test DP using ranges of suitable $f$-divergences \citep{asoodeh2021threevariants}, while Asoodeh, Aliakbarpour, and Calmon characterize LDP through contraction of the $E_\gamma$ divergence \citep{asoodeh2021ldpcontraction}. Sadeghi, Ding, and Rakotoarivelo study information privacy through log-lift and relaxed log-lift notions \citep{sadeghi2020privacywatchdog}. These works motivate the divergence-based accounting viewpoint used here, but our emphasis is on how support restrictions change the exact channel-level defect and the resulting learning risk.

Finite-support and integer-valued noise mechanisms for count queries motivate bounded perturbation supports \citep{sadeghi2020countqueries,sadeghi2024abs}, while Skellam and Poisson-binomial mechanisms show the importance of discrete noise laws in private distributed learning \citep{agarwal2021skellam,chen2022poissonbinomial}. Mechanism-level sparsity is also connected to information-geometric privacy design, where local approximations explain privacy--utility tradeoffs \citep{zamani2025infogeometric}. In contrast, the present paper uses exact finite-alphabet privacy defects and exact finite-sample risk formulas to design support-constrained randomizers for robust private learning with explicitly delimited context adaptation.

\subsection{Contributions}

The main contributions are as follows.
\begin{enumerate}[leftmargin=*,label=(\roman*)]
\item We formulate support-constrained local randomization under range-restricted approximate LDP and decompose its exact hockey-stick defect into support leakage and overlap excess loss. For line-supported sparse randomized response, we derive the exact pairwise and range-$H$ defects, solve the joint design over $(p,q,t)$ in closed form at each radius, and prove the support-size lower bound $1+H/\delta-e^\varepsilon$ in the strong approximate-privacy regime.
\item We characterize private distribution estimation through an exact finite-sample inverse risk. For every invertible finite channel, worst-case risk over a closed convex distribution set reduces to a Euclidean projection; on the simplex, the least-favorable law has an explicit water-filling form. We also give distribution-shift stability and exact ranking-reversal criteria.
\item We derive the exact privacy defect of cyclic sparse randomized response at every admissible radius and prove a cyclic minimax-robustness theorem. The uniform law is uniquely least favorable, nominal and minimax channel rankings coincide, and Fourier eigenvalues give the exact minimax risk together with spectral lower and upper bounds.
\item We develop distributionally robust and privacy-range-robust calibration. A boundary-normalized nonhomogeneous path family exhibits a strict nominal--minimax tradeoff: over a disclosed deterministic grid, robust calibration changes the privacy-matched mechanism and reduces worst-case risk by about $45\%$. We also state precise independence and composition conditions for public-context and history-dependent selection.
\item We evaluate all claims with deterministic exact privacy and risk calculations, matched-guarantee baselines, and a supplied reproduction script. The experiments cover support scaling, finite-sample risk, cyclic distribution shift, the nonhomogeneous robust example, privacy-range uncertainty, and public-context selection. We finally explain how the support-leakage design principle extends to sparse Laplace, Gaussian, staircase, and score-based channels without attributing the randomized-response optimality results to those families.
\end{enumerate}

\section{Privacy Model and Support Constraints}

Let $\mathcal{X}$ and $\mathcal{Y}$ be discrete input and output alphabets,
each finite or countably infinite. The privacy definitions and the
support-mismatch decomposition below do not require
$\mathcal{X}=\mathcal{Y}$. We use three explicit specializations. Sections
III--IV analyze the translation-invariant line model
$\mathcal{X}=\mathcal{Y}=\mathbb Z$. Section V first treats an arbitrary
finite square channel and then specializes to the cycle
$\mathcal{X}=\mathcal{Y}=\mathbb Z_K$. Section VI-B uses the finite path
$\mathcal{X}=\mathcal{Y}=\{0,\ldots,K-1\}$, where boundary normalization is
input dependent. Equality of the alphabets is thus a same-domain reporting
assumption used by our mechanisms, not a requirement of local privacy; the
finite square assumption is needed only for inverse-channel estimation. A
local randomizer is a channel $M(\cdot\mid x)$ from inputs in
$\mathcal{X}$ to outputs in $\mathcal{Y}$. It is pure $\varepsilon$-LDP if for
all $x,x'\in\X$ and all $A\subseteq\Y$,
\begin{equation}
M(A\mid x)\le e^\varepsilon M(A\mid x'),
\end{equation}
and it is $(\varepsilon,\delta)$-LDP if
\begin{equation}
M(A\mid x)\le e^\varepsilon M(A\mid x')+\delta.
\end{equation}
Because $\Y$ is discrete, the approximate condition is equivalent to the hockey-stick, or $E_\gamma$, divergence bound with $\gamma=e^\varepsilon$~\citep{dwork2014book,asoodeh2021ldpcontraction}:
\begin{equation}\label{eq:hockey-aistats}
\Delta^M_\varepsilon(x,x'):=\sum_{y\in\Y}\pos{M(y\mid x)-e^\varepsilon M(y\mid x')}\le\delta.
\end{equation}
We use a relation-restricted version of the definition. For a symmetric relation $\mathcal R\subseteq\X\times\X$, a channel is $(\varepsilon,\delta)$-LDP on $\mathcal R$ if \eqref{eq:hockey-aistats} holds for all $(x,x')\in\mathcal R$. On the integer line, for an integer $H\ge 1$, define
\begin{equation}
\mathcal R_H=\{(x,x'):1\le |x-x'|\le H\}.
\end{equation}
We call this $(\varepsilon,\delta)$-LDP on range $H$. It is a relational local-privacy guarantee, not ordinary all-pairs LDP unless $\mathcal R_H=\X\times\X$. The usual all-pairs definition corresponds to $\mathcal R=\X\times\X$. We state the protected relation and the attained defect explicitly in every experiment.

The support-mismatch identity below is the accounting tool used in the paper. If $S(x)=\operatorname{supp}M(\cdot\mid x)$, then
\begin{align}\label{eq:decomp-aistats}
\Delta^M_\varepsilon(x,x')
=M(S(x)\setminus S(x')\mid x)
+
\sum_{y\in S(x)\cap S(x')}
\pos{M(y\mid x)-e^\varepsilon M(y\mid x')}.
\end{align}
The first term is support leakage: outputs possible under $x$ but impossible under $x'$. The second term is overlap excess loss: outputs possible under both inputs but whose likelihood ratio exceeds $e^\varepsilon$.
Pure local privacy cannot tolerate an output that is possible under one protected input and impossible under another. Thus, before considering approximate guarantees, it is useful to isolate the common-support requirement that pure privacy imposes. The next result formalizes this obstruction and explains why finite-radius mechanisms on an unbounded line necessarily require either approximation or a restricted privacy relation.
\begin{proposition}\label{prop:pure-obstruction}
If $M$ has finite privacy parameter $\varepsilon$ under pure LDP on a relation $\mathcal R$, then $S(x)=S(x')$ for every $(x,x')\in\mathcal R$. In particular, a finite-radius line mechanism cannot satisfy pure all-pairs LDP on an unbounded integer alphabet.
\end{proposition}

\par\noindent\emph{Proof.} The proof is deferred to Appendix A.

We consider support-constrained designs of the form
\begin{equation}\label{eq:general-design}
\min_M \; \E[c(X,Y)]
\quad\text{subject to}\quad
\operatorname{supp}M(\cdot\mid x)\subseteq S_t(x),
\end{equation}
where $X$ is the private input, $Y$ is the reported output, and
$c:\X\times\Y\to[0,\infty)$ is a user-specified reporting cost or
statistical distortion. This is a cost-minimization formulation rather than an
abstract utility-maximization problem; equivalently, one could take a utility
$u(x,y)=-c(x,y)$. Throughout, we assume that $c(x,y)\ge 0$, that
$c(x,x)=0$, and that $c(x,y)<\infty$ on the admissible support. The constraint
is imposed together with $(\varepsilon,\delta)$-LDP on $\mathcal R_H$.
The radius $t$ is part of the design. The next sections solve this problem for
sparse randomized response with the line-metric cost $c(x,y)=|x-y|$, so the
objective becomes the first reporting-distance moment.
Support restriction by itself does not imply a communication reduction: a
decoder can encode a local offset using fewer symbols only when it possesses a
public anchor or other side information that identifies the relevant local
support.

\section{Sparse Randomized Response}

Throughout Sections III--IV, the alphabet is the unbounded integer line
$\X=\Y=\mathbb Z$ and the channel is translation invariant. Finite cyclic and
finite-path versions are introduced separately in Sections V and VI.
For an integer support radius $t\ge 1$, let
$S_t(x)=\{x-t,\ldots,x+t\},\ s=2t+1.$
The sparse randomized-response mechanism is
\begin{equation}\label{eq:sparse-rr}
M^{\RR}(y\mid x)=
\begin{cases}
p, & y=x,\\
q, & y\in S_t(x)\setminus\{x\},\\
0, & y\notin S_t(x),
\end{cases}
\end{equation}
where $p\ge q\ge0$ and $p+2tq=1$. The condition $p\ge q$ encodes the usual truthful-bias interpretation of randomized response.

For a pair of inputs, the privacy loss is determined by how their radius-$t$ supports overlap. Disjoint supports reveal the input region completely, partial overlap can place the truthful output in the support-mismatch set, and strong overlap leaves only boundary leakage together with a possible truthful-symbol likelihood-ratio excess. The next result gives the exact defect in these three geometric regimes.
\begin{proposition}\label{prop:rr-defect-aistats}
For inputs separated by $h=|x-x'|$, the ordered defect of \eqref{eq:sparse-rr} is
\begin{equation}\label{eq:rr-pair-defect}
\Delta^{\RR}_\varepsilon(x,x')=
\begin{cases}
1, & h>2t,\\
hq+\pos{p-e^\varepsilon q}, & 1\le h\le t,\\
p+q(h-1), & t<h\le2t.
\end{cases}
\end{equation}
\end{proposition}

\par\noindent\emph{Proof.} The proof is deferred to Appendix A.

The pairwise formula can be maximized analytically over every protected separation. In each feasible overlap regime the largest defect occurs at an endpoint of the protected range, so no numerical maximization over input pairs is needed. The resulting expression below is the exact privacy constraint used throughout the design problem.
\begin{corollary}\label{cor:range-defect}
The exact range-$H$ defect of sparse randomized response is
\begin{equation}\label{eq:rr-range-defect}
\delta_t(p,q;\varepsilon,H)=
\begin{cases}
1, & H>2t,\\
Hq+\pos{p-e^\varepsilon q}, & H\le t,\\
1-(2t-H+1)q, & t<H\le2t.
\end{cases}
\end{equation}
\end{corollary}

\par\noindent\emph{Proof.} The proof is deferred to Appendix A.

The first reporting-distance moment is
\begin{equation}\label{eq:rr-distortion}
R_1^{\RR}(p,q,t):=\sum_y |y-x|M^{\RR}(y\mid x)=q t(t+1).
\end{equation}
For fixed $t$, distortion is increasing in $q$: more probability on the nontruthful support points produces larger expected reporting distance.

\section{Joint Privacy-Distortion Optimization}

We now optimize \eqref{eq:rr-distortion} subject to the exact range-$H$ privacy constraint in \eqref{eq:rr-range-defect}.
At a fixed radius, the reporting-distance objective increases with the off-truth probability $q$. The optimal channel therefore uses the smallest $q$ that reaches the target privacy level, provided that the radius can attain that level at all. The next theorem identifies this boundary point in closed form for both overlap regimes and leaves only a discrete search over the radius.
\begin{theorem}\label{thm:joint-aistats}
Fix $\varepsilon>0$, $\delta\in(0,1)$, and an integer range $H\ge1$. For a fixed radius $t$, the distortion-minimizing feasible value of $q$ is
\begin{equation}\label{eq:qstar-aistats}
q_t^*(\delta)=
\begin{cases}
\dfrac{1-\delta}{2t+e^\varepsilon-H}, & H\le t\text{ and }\delta\ge \dfrac{H}{e^\varepsilon+2t},\\[1em]
\dfrac{1-\delta}{2t-H+1}, & t<H\le2t\text{ and }\delta\ge \dfrac{H}{2t+1},\\[1em]
\text{infeasible}, & \text{otherwise},
\end{cases}
\end{equation}
with $p_t^*(\delta)=1-2tq_t^*(\delta)$. Hence the globally optimal support radius and reporting probabilities are obtained by the one-dimensional search
\begin{equation}\label{eq:global-search-aistats}
t^*\in \arg\min_{t:q_t^*(\delta)\text{ feasible}} q_t^*(\delta)t(t+1),
\qquad
p^*=1-2t^*q_{t^*}^*(\delta).
\end{equation}
The feasible set is nonempty and the minimum in
\eqref{eq:global-search-aistats} is attained.
\end{theorem}

\par\noindent\emph{Proof.} The proof is deferred to Appendix A.

Before optimizing distortion, one must know whether a chosen support radius can satisfy the desired privacy budget. Minimizing the exact defect over the truthful-bias interval produces a radius-dependent privacy floor. The following result gives that floor and therefore characterizes feasibility exactly.
\begin{corollary}\label{cor:privacy-floors}
For radius $t$, the minimum achievable range-$H$ defect over all truthful probabilities $p,q$ is
\begin{equation}\label{eq:privacy-floor}
\delta_t^{\min}(\varepsilon,H)=
\begin{cases}
1, & H>2t,\\
\dfrac{H}{e^\varepsilon+2t}, & H\le t,\\[0.8em]
\dfrac{H}{2t+1}, & t<H\le2t.
\end{cases}
\end{equation}
Thus a target $\delta$ is feasible at radius $t$ if and only if $\delta\ge \delta_t^{\min}(\varepsilon,H)$.
\end{corollary}

\par\noindent\emph{Proof.} The proof is deferred to Appendix A.

The privacy floor also limits how aggressively the output support can be truncated. In the strong approximate-privacy regime, every radius smaller than the protected range lies above the target defect, and the remaining branch can be inverted to obtain an explicit minimum support size. This yields the following converse.
\begin{corollary}\label{cor:support-lower-bound}
If
$\delta<\frac{H}{2H-1},$
then every feasible radius satisfies $t\ge H$, and its support size
$s=2t+1$ obeys
$s\ge 1+\frac{H}{\delta}-e^\varepsilon.$
Consequently, for fixed $(H,\varepsilon)$, strong approximate privacy forces
$s=\Omega(H/\delta)$.
\end{corollary}

\par\noindent\emph{Proof.} The proof is deferred to Appendix A.

The following remark explains why the common-support randomized-response calibration is not generally optimal in the approximate range-restricted setting.
\begin{remark}
The common-support randomized-response calibration sets $p=e^\varepsilon q$. Theorem~\ref{thm:joint-aistats} shows that this calibration is not generally optimal for approximate range-$H$ privacy with shifting sparse supports. For looser target $\delta$, the optimized mechanism can use smaller off-truth probability $q$ and larger truthful probability $p$, reducing reporting distortion while preserving the prescribed range-$H$ guarantee.
\end{remark}

The theorem has two useful interpretations. First, feasibility is governed by a privacy floor: at a fixed support radius, no choice of probabilities can push the defect below \eqref{eq:privacy-floor}. Second, once feasibility holds, the optimal probabilities are explicit. The remaining search is discrete and one-dimensional.

\subsection{Risk-Aware Calibration}

Theorem~\ref{thm:joint-aistats} optimizes reporting distance. For statistical estimation, however, the conditioning of the channel matrix also matters. The same exact privacy formula yields a risk-aware design problem.
For a fixed radius $t$, define the feasible interval
\begin{equation}\label{eq:q-interval}
\mathcal Q_t(\delta)=\{q\in[0,1/(2t+1)]:\delta_t(1-2tq,q;\varepsilon,H)\le\delta\}.
\end{equation}
When $H\le t$ and the target is feasible, this interval is
\begin{equation}\label{eq:q-interval-case1}
\mathcal Q_t(\delta)=\left[\frac{1-\delta}{2t+e^\varepsilon-H},\;\min\left\{\frac{\delta}{H},\frac{1}{2t+1}\right\}\right],
\end{equation}
and when $t<H\le2t$ it is
\begin{equation}\label{eq:q-interval-case2}
\mathcal Q_t(\delta)=\left[\frac{1-\delta}{2t-H+1},\;\frac{1}{2t+1}\right].
\end{equation}
If $H>2t$ or the corresponding privacy floor in Corollary~\ref{cor:privacy-floors} exceeds $\delta$, then $\mathcal Q_t(\delta)$ is empty.
Reporting distance alone does not control estimation quality because inverse estimation can amplify noise when the channel is poorly conditioned. Once the exact privacy constraint is expressed as an interval of admissible $q$ values, however, the statistical design still has only one continuous degree of freedom at each radius. The next result makes this reduction explicit and excludes singular channels whose inverse-estimation risk is unbounded.
\begin{proposition}\label{prop:risk-aware}
Fix a finite square alphabet, a finite admissible set of radii $\mathcal T$, a
design prior $\pi_0$, and a sample size $n$. For sparse randomized-response
candidates whose privacy sublevel set is $\mathcal Q_t(\delta)$, the exact-risk
design problem
\begin{equation}\label{eq:risk-aware-problem}
\min_{t\in\mathcal T,p,q}\; \frac{1}{n}\tr\!\left(P^{-T}\!\left(\diag(\pi_0P)-(\pi_0P)^T(\pi_0P)\right)P^{-1}\right)
\end{equation}
subject to range-$H$ $(\varepsilon,\delta)$-LDP, $p=1-2tq$, and $p\ge q$, reduces to the finite collection of one-dimensional problems
\begin{equation}\label{eq:risk-aware-search}
\min_{\substack{t\in\mathcal T:\mathcal Q_t(\delta)\ne\emptyset}}\;\min_{q\in\mathcal Q_t(\delta)} R_{\mathrm{est}}(t,q),
\end{equation}
where $R_{\mathrm{est}}(t,q)$ denotes the exact risk in \eqref{eq:risk-aware-problem} for an invertible channel with radius $t$ and off-truth probability $q$; singular candidates have infinite inverse-estimation risk and are excluded.
Under the line-overlap geometry, $\mathcal Q_t(\delta)$ is given by
\eqref{eq:q-interval-case1}--\eqref{eq:q-interval-case2}. For a finite cycle
outside that regime, the same reduction holds after replacing it by the
sublevel set of the exact cyclic defect in Proposition~\ref{prop:cyclic-defect}.
\end{proposition}

\par\noindent\emph{Proof.} The proof is deferred to Appendix A.

The risk-aware problem uses the same privacy characterization but a different utility criterion. It can choose a larger support radius than the distortion-optimal design if doing so improves the channel spectrum and reduces inverse-estimation variance. Equations~\eqref{eq:q-interval-case1}--\eqref{eq:q-interval-case2} use the line-overlap geometry. On a finite cycle whose support nearly fills the alphabet, the exact cyclic defect derived below must be used instead.

\section{Distribution Estimation}
We now connect the optimized mechanism to a downstream statistical task. Let $X_1,\ldots,X_n\sim\pi$ be i.i.d. private samples on a finite alphabet of size $K$. Each user reports $Y_i\sim M(\cdot\mid X_i)$. Let $P$ be the channel matrix with entries $P_{xy}=M(y\mid x)$, and let $\hat m$ be the empirical distribution of the reported symbols.
From this point onward, we assume
$\mathcal{X}=\mathcal{Y}$ with $|\mathcal{X}|=K$, so that the channel matrix
$P$ is square. Invertibility is required only for the linear
inverse-estimation results.
The privatized histogram is multinomial with mean $m=\pi P$. Applying the inverse channel transforms both its mean and covariance, so the estimation error can be computed without an asymptotic approximation. The following theorem gives this exact finite-sample risk and fixes the row-vector outer-product convention used below.
\begin{theorem}\label{thm:linear-risk}
Assume $P$ is invertible and define the linear inverse estimator
\begin{equation}
\hat\pi=\hat m P^{-1}.
\end{equation}
Let $m=\pi P$ be the output distribution. Then $\hat\pi$ is unbiased and
\begin{equation}\label{eq:exact-risk}
\E\|\hat\pi-\pi\|_2^2
=\frac{1}{n}\tr\!\left(P^{-T}(\diag(m)-m^Tm)P^{-1}\right),
\end{equation}
where $m^Tm$ is the outer product under the row-vector convention.
\end{theorem}

\par\noindent\emph{Proof.} The proof is deferred to Appendix A.

The exact expression \eqref{eq:exact-risk} shows that utility depends not only on reporting distortion, but also on the conditioning of the channel matrix. More importantly, its dependence on the unknown input distribution is a concave quadratic. Completing the square converts worst-case distribution selection over any closed convex uncertainty set into a Euclidean projection, as stated next.
For a nonempty closed convex set $\Pi\subseteq\mathbb{R}^K$ and
$z\in\mathbb{R}^K$, define the Euclidean distance to $\Pi$ and the Euclidean
projection onto $\Pi$ by
$
\operatorname{dist}(z,\Pi)
:=
\inf_{u\in\Pi}\|z-u\|_2,
\
\operatorname{Proj}_{\Pi}(z)
:=
\arg\min_{u\in\Pi}\|z-u\|_2.$
Because $\Pi$ is nonempty, closed, and convex, the projection exists and is
unique.

\begin{theorem}\label{thm:projection-risk}
Let
\begin{equation}\label{eq:risk-coefficients}
B=P^{-1}P^{-T},\qquad
d=(B_{11},\ldots,B_{KK})^T,\qquad
a(P)=Pd.
\end{equation}
Then the exact risk can be written as
\begin{align}
R(\pi;P)
:=\E_\pi\|\hat\pi-\pi\|_2^2=\frac{1}{n}\left(\pi a(P)-\|\pi\|_2^2\right)
=\frac{1}{n}\left[
\frac{\|a(P)\|_2^2}{4}
-\left\|\pi^T-\frac{a(P)}{2}\right\|_2^2
\right].\label{eq:quadratic-risk}
\end{align}
For any nonempty closed convex uncertainty set $\Pi\subseteq\Delta_K$, the
least-favorable law is unique and satisfies
\begin{equation}\label{eq:projection-law}
(\pi_\Pi^*(P))^T
=\operatorname{Proj}_{\Pi}\!\left(\frac{a(P)}{2}\right).
\end{equation}
The corresponding exact worst-case risk is
\begin{equation}\label{eq:projection-value}
\sup_{\pi\in\Pi}R(\pi;P)
=\frac{1}{n}\left[
\frac{\|a(P)\|_2^2}{4}
-\operatorname{dist}^2\!\left(\frac{a(P)}{2},\Pi\right)
\right].
\end{equation}
For the full simplex $\Delta_K$, the projection has the water-filling form
\begin{equation}\label{eq:waterfilling}
\pi_i^*=\max\left\{\frac{a_i(P)}{2}-\tau,0\right\},
\qquad \sum_{i=1}^K\pi_i^*=1,
\end{equation}
where the scalar $\tau$ is uniquely determined by normalization.
\end{theorem}

\par\noindent\emph{Proof.} The proof is deferred to Appendix A.

Theorem~\ref{thm:projection-risk} removes the inner distributional optimization
from robust channel design. For a family $\mathcal P$ of privacy-feasible
invertible channels, $\min_{P\in\mathcal P}\sup_{\pi\in\Pi}R(\pi;P)$ requires
one Euclidean projection per candidate channel.
The quadratic risk representation also quantifies sensitivity to deployment shift without recalculating the channel covariance from scratch. A direct Lipschitz bound in total variation follows by separating the linear channel term from the common quadratic term.
\begin{corollary}\label{cor:risk-stability}
For any two input laws $\pi$ and $\pi_0$,
\begin{equation}\label{eq:risk-stability}
|R(\pi;P)-R(\pi_0;P)|
\le \frac{\|a(P)\|_\infty+2}{n}\,\|\pi-\pi_0\|_1.
\end{equation}
Consequently, over a total-variation ball of radius $\rho$ centered at
$\pi_0$, the risk differs from $R(\pi_0;P)$ by at most
$2\rho(\|a(P)\|_\infty+2)/n$.
\end{corollary}

\par\noindent\emph{Proof.} The proof is deferred to Appendix A.

When two channels are compared, their common term $-\|\pi\|_2^2/n$ cancels. Their risk difference is therefore linear in the deployment distribution, and its extreme values occur at alphabet vertices. The next result gives an exact criterion for whether distribution shift can reverse the ranking of two mechanisms.
\begin{corollary}\label{cor:ranking-shift}
For two invertible channels $P_1$ and $P_2$, let
$v=a(P_1)-a(P_2)$. Then
\begin{equation}\label{eq:ranking-shift}
R(\pi;P_1)-R(\pi;P_2)=\frac{1}{n}\pi v.
\end{equation}
Over the full simplex, the risk difference ranges from
$\min_i v_i/n$ to $\max_i v_i/n$. Hence a strict ranking reversal under
distribution shift is possible if and only if
$\min_i v_i<0<\max_i v_i$. In particular, a constant $v$ makes the ranking
independent of the input distribution.
\end{corollary}

\par\noindent\emph{Proof.} The result follows by subtracting
\eqref{eq:quadratic-risk} for the two channels; a linear functional on the
simplex attains its extrema at simplex vertices.

For cyclic ordered alphabets, both privacy and conditioning can be written
explicitly. Let $d_K(x,x')$ denote cyclic distance and let
$1\le h\le\lfloor K/2\rfloor$.
On a finite cycle, supports can wrap around and overlap again when the radius is large. The line-based mismatch count then overstates leakage, especially near full support. The following result replaces it with the exact cyclic mismatch cardinality and gives a defect formula valid at every radius below $K/2$.
\begin{proposition}\label{prop:cyclic-defect}
For cyclic sparse randomized response with radius $t<K/2$, set
\begin{equation}\label{eq:cyclic-mismatch}
r_{K,t}(h)=\min\{h,\,2t+1,\,K-2t-1\}.
\end{equation}
The exact ordered defect between inputs at cyclic distance $h$ is
\begin{align}
\Delta_{\varepsilon,K}^{\RR}(h)
=q\,r_{K,t}(h)
+\1\{h\le t\}\pos{p-e^\varepsilon q}+\1\{h>t\}(p-q).\label{eq:cyclic-defect}
\end{align}
Hence the exact cyclic range-$H$ defect is the maximum of
\eqref{eq:cyclic-defect} over
$1\le h\le\min\{H,\lfloor K/2\rfloor\}$. If $2t+H<K$, this reduces to the
line formula in Corollary~\ref{cor:range-defect}. At full support on an odd
cycle, $r_{K,t}(h)=0$ and the defect is simply
$\pos{p-e^\varepsilon q}$.
\end{proposition}

\par\noindent\emph{Proof.} The proof is deferred to Appendix A.

Translation invariance on the cycle makes the channel circulant. Consequently, invertibility and noise amplification are governed by the discrete Fourier transform of one channel row. The next proposition records the eigenvalues that are used in the minimax risk analysis.
\begin{proposition}\label{prop:cyclic-spectrum}
Let the alphabet be $\Z_K$ and let $S_t(x)=\{x-t,\ldots,x+t\}$ modulo $K$, with $t<K/2$. The sparse randomized-response channel is circulant with Fourier eigenvalues
\begin{equation}\label{eq:fourier-eigs}
\lambda_\ell=p+2q\sum_{j=1}^t \cos\left(\frac{2\pi \ell j}{K}\right),
\qquad \ell=0,\ldots,K-1.
\end{equation}
The channel is invertible if and only if $\min_\ell |\lambda_\ell|>0$.
\end{proposition}

\par\noindent\emph{Proof.} The proof is deferred to Appendix A.

For a circulant channel, every diagonal entry of $P^{-1}P^{-T}$ is identical. This symmetry makes the channel-dependent part of the exact risk constant across input distributions, leaving the same negative quadratic term for every candidate. The next theorem converts that observation into an exact minimax characterization and shows that distributional robustness cannot change rankings inside a common feasible circulant family.
\begin{theorem}\label{thm:cyclic-minimax}
For any invertible circulant stochastic channel $P$, let
$\lambda_0=1,\lambda_1,\ldots,\lambda_{K-1}$ be its Fourier eigenvalues and
define
\begin{equation}\label{eq:cyclic-cp}
c(P)=\frac{1}{K}\sum_{\ell=0}^{K-1}|\lambda_\ell|^{-2}.
\end{equation}
For every input distribution $\pi$,
\begin{equation}\label{eq:cyclic-exact-risk}
R(\pi;P)=\frac{1}{n}\left[c(P)-\|\pi\|_2^2\right].
\end{equation}
The uniform distribution is the unique least-favorable law on $\Delta_K$, and
\begin{equation}\label{eq:uniform-risk}
\sup_{\pi\in\Delta_K}R(\pi;P)
=\frac{1}{nK}\sum_{\ell=1}^{K-1}\frac{1}{|\lambda_\ell|^2}.
\end{equation}
Moreover, for any two invertible circulant channels $P_1,P_2$, the sign of
$R(\pi;P_1)-R(\pi;P_2)$ is independent of $\pi$.
\end{theorem}

\par\noindent\emph{Proof.} The proof is deferred to Appendix A.

Thus nominal-risk and minimax-risk calibration over the same feasible cyclic
family must select the same channel (up to ties). A claimed prior-dependent
cyclic robust choice signals unequal feasibility constraints, numerical error,
or a noncirculant implementation.
The exact minimax expression involves the reciprocal energy of every nonconstant Fourier mode. Cauchy--Schwarz relates this quantity to the total nonconstant spectral energy, while the smallest nonconstant eigenvalue controls it from above. The resulting two-sided bound separates aggregate channel energy from worst-mode conditioning.
\begin{corollary}\label{cor:spectral-sandwich}
For cyclic sparse randomized response, define
\begin{equation}\label{eq:spectral-energy}
E(P)=K(p^2+2tq^2)-1
=\sum_{\ell=1}^{K-1}|\lambda_\ell|^2,
\qquad
\mu(P)=\min_{1\le\ell\le K-1}|\lambda_\ell|.
\end{equation}
Then, we have
\begin{equation}\label{eq:spectral-sandwich}
\frac{(K-1)^2}{nK\,E(P)}
\le \sup_{\pi\in\Delta_K}R(\pi;P)
\le \frac{K-1}{nK\,\mu(P)^2}.
\end{equation}
In particular, the earlier general bound
$R(\pi;P)\le1/(n\mu(P)^2)$ remains valid but is looser.
\end{corollary}

\par\noindent\emph{Proof.} The proof is deferred to Appendix A.

For full $K$-ary randomized response with $p=e^\varepsilon/(e^\varepsilon+K-1)$ and $q=1/(e^\varepsilon+K-1)$, the nonzero-frequency eigenvalue is $p-q=(e^\varepsilon-1)/(e^\varepsilon+K-1)$. Thus the familiar full-support mechanism is always invertible, but its inverse can be poorly conditioned when $K$ is large and $\varepsilon$ is small. A sparse channel can have a much larger truthful probability, but its spectrum depends on the chosen support radius. This is why the experiments below report both privacy feasibility and estimation risk.

\section{Robust Private Learning and Safe Adaptation}\label{sec:robust-adaptive}

The preceding results give exact privacy and estimation-risk formulas for a fixed support radius and a fixed design distribution. In deployment, however, the private data distribution, the relevant privacy range, or the public application domain may change after a mechanism has been calibrated. This section uses the exact risk and defect expressions above to turn those changes into finite-dimensional design problems and states precisely when channel selection from context is privacy safe.

\subsection{Distributionally robust calibration}

Let $\Pi$ be an uncertainty set of possible input distributions. For example, $\Pi$ may be a finite collection of environments, a neighborhood of a nominal distribution, or a set of plausible distributions estimated from public data. On a finite alphabet, let $\mathcal T$ denote the finite set of admissible support radii. For a radius $t$ and off-truth probability $q$, let $R_{\mathrm{est}}(t,q;\pi)$ be the exact inverse-estimation risk in Theorem~\ref{thm:linear-risk}. The robust calibration problem is
\begin{equation}\label{eq:jsait-robust-risk}
\min_{\substack{t\in\mathcal T:\mathcal Q_t(\delta)\ne\emptyset}}\;\min_{q\in\mathcal Q_t(\delta)}\;\sup_{\pi\in\Pi}R_{\mathrm{est}}(t,q;\pi).
\end{equation}
This problem keeps the same privacy feasibility set as Proposition~\ref{prop:risk-aware}, but replaces the nominal risk by worst-case risk over the uncertainty set.

For a finite set of possible deployment environments, robustness adds only a maximum over finitely many exact risks. Privacy feasibility and the one-dimensional channel parameterization remain unchanged. The following result shows that the full minimax problem is therefore still a finite radius search with a scalar inner optimization.
\begin{proposition}\label{prop:finite-robust-calibration}
Suppose $\Pi=\{\pi^{(1)},\ldots,\pi^{(L)}\}$ is finite. Then distributionally robust sparse randomized-response calibration reduces to a finite search over $\mathcal T$ and a one-dimensional optimization over $q$:
\begin{equation}\label{eq:finite-robust-search}
\min_{\substack{t\in\mathcal T:\mathcal Q_t(\delta)\ne\emptyset}}\;\min_{q\in\mathcal Q_t(\delta)}\;\max_{1\le \ell\le L}R_{\mathrm{est}}(t,q;\pi^{(\ell)}).
\end{equation}
For each candidate pair $(t,q)$, the objective is evaluated exactly from \eqref{eq:exact-risk}.
\end{proposition}
\begin{proof}
For each fixed radius $t$, Proposition~\ref{prop:risk-aware} gives the feasible interval $\mathcal Q_t(\delta)$ for the off-truth probability. For each $q$ in this interval and each $\pi^{(\ell)}$, Theorem~\ref{thm:linear-risk} gives the exact finite-sample risk. Taking the maximum over the finite set $\Pi$ yields the robust objective, and searching over the finite set of feasible radii gives \eqref{eq:finite-robust-search}.
\end{proof}

This formulation is useful when the learner is trained or calibrated under one distribution but deployed under another. It also highlights why distortion-optimality is not necessarily robust: a support radius that minimizes reporting distance can make the inverse channel poorly conditioned for some distributions, increasing worst-case estimation risk.

For a closed convex uncertainty set, Theorem~\ref{thm:projection-risk} is
stronger than a finite enumeration: the inner supremum in
\eqref{eq:jsait-robust-risk} is the projection value in
\eqref{eq:projection-value}. On a cycle, however,
Theorem~\ref{thm:cyclic-minimax} shows that every candidate's dependence on
$\pi$ contains the same term $-\|\pi\|_2^2/n$. Nominal and robust calibration
therefore cannot select different cyclic channels when the feasible family is
the same. A nonhomogeneous model is required for a genuine mechanism change.

\subsection{Nonhomogeneous path calibration}

Let $\X=\Y=\{0,\ldots,K-1\}$ be a finite path. We consider the
boundary-normalized score channel
\begin{equation}\label{eq:path-channel}
P_{\theta,t}(y\mid x)=
\frac{
\exp\{-\eta|y-x|+\beta(y-x)\}
b^{\1\{y=x\}}\1\{|y-x|\le t\}}
{Z_x(\theta,t)},
\end{equation}
where $\theta=(\eta,b,\beta)$, $\eta\ge0$, $b\ge1$, $\beta\in\R$, and
$Z_x(\theta,t)$ is the row normalizer over the truncated path. Near a
boundary, $Z_x$ depends on $x$; when $\beta\ne0$, the kernel is also
directionally asymmetric. Thus the channel is generally neither circulant nor
doubly stochastic.

For every candidate, its exact relational defect is the finite maximum
\begin{equation}\label{eq:path-defect}
\delta_H(P_{\theta,t})=
\max_{1\le |x-x'|\le H}
\sum_{y=0}^{K-1}
\pos{P_{\theta,t}(y\mid x)
-e^\varepsilon P_{\theta,t}(y\mid x')}.
\end{equation}
For a closed convex uncertainty set $\Pi$, exact robust path calibration is
therefore
\begin{equation}\label{eq:path-robust-program}
\min_{\substack{t,\theta:\ \delta_H(P_{\theta,t})\le\delta\\
P_{\theta,t}\ \mathrm{invertible}}}
\frac{1}{n}\left[
\frac{\|a(P_{\theta,t})\|_2^2}{4}
-\operatorname{dist}^2\!\left(\frac{a(P_{\theta,t})}{2},\Pi\right)
\right].
\end{equation}
For the nonhomogeneous path family, the projection formula evaluates the worst input law exactly for every candidate channel. Thus a disclosed finite candidate grid can be optimized globally by exhaustive screening, while a compact continuous parameter set admits an optimizer once singular channels are uniformly excluded. The next proposition states both guarantees.
\begin{proposition}\label{prop:path-robust}
For any finite candidate family of path channels, screening candidates with
\eqref{eq:path-defect}, discarding singular matrices, and minimizing the
projection value in \eqref{eq:path-robust-program} returns the exact global
minimax solution within that family. More generally, if the parameter set is
compact, the channel map is continuous, and feasible candidates satisfy
$\sigma_{\min}(P_{\theta,t})\ge\kappa>0$, then a robust minimizer exists.
\end{proposition}
\begin{proof}
The finite statement follows by exact evaluation of every feasible candidate.
For the continuous statement, the defect is a maximum of finitely many
continuous functions, so the privacy-feasible set is closed. The singular-value
condition makes inversion and the objective in \eqref{eq:path-robust-program}
continuous. A continuous function attains its minimum on the resulting compact
feasible set.
\end{proof}

Corollary~\ref{cor:ranking-shift} supplies an analytical diagnostic for this
family: two path mechanisms can exchange their ranking precisely when the
coordinates of $a(P_1)-a(P_2)$ straddle zero. The experiment below verifies
this condition for the reported nominal and minimax designs.

\subsection{Robustness to privacy-range mismatch}

The range parameter $H$ specifies which input pairs are protected. In deployment, the relevant range may be larger than the nominal design range. The exact defect formula makes this mismatch explicit. If a mechanism is calibrated for $H_0$ but deployed under $H_1>H_0$, its actual privacy defect is $\delta_t(p,q;\varepsilon,H_1)$, which can exceed the target.
The protected pair sets are nested in the range parameter. A certificate at the largest anticipated range therefore covers every smaller range without additional privacy loss. The following corollary states this monotonicity-based robust design rule.
\begin{corollary}\label{cor:range-robust}
If a sparse randomized-response mechanism satisfies $\delta_t(p,q;\varepsilon,H_{\max})\le\delta$, then it satisfies $(\varepsilon,\delta)$-LDP on every range $H\le H_{\max}$. Equivalently, calibrating at the largest anticipated range protects all smaller ranges.
\end{corollary}
\begin{proof}
The range-$H$ defect is the maximum of $\Delta_\varepsilon^{\RR}(0,h)$ over $1\le h\le H$. If $H\le H_{\max}$, this maximum is taken over a subset of the separations used to define $\delta_t(p,q;\varepsilon,H_{\max})$. Hence it cannot be larger.
\end{proof}

This result is intentionally conservative. When the deployment range is known, one should calibrate at that range. When the range is uncertain, $H_{\max}$ gives a robust certificate.

\subsection{Public-context adaptation}

Adaptive systems often have access to public context before choosing a local randomizer: a domain label, a public estimate of the alphabet geometry, a public accuracy requirement, or a privacy range agreed upon for the current task. Selecting among already certified channels preserves privacy only when the selector is independent of the current protected record or is itself privatized and composed with the selected channel. The following proposition distinguishes conditional guarantees, joint guarantees, private selectors, fresh users, and repeated participation.

\begin{proposition}\label{prop:public-context-adaptation}
Let $C$ be observed before applying a channel, and suppose every $M_c$ is
$(\varepsilon,\delta)$-LDP on a common relation $\mathcal R$.
\begin{enumerate}[leftmargin=*,label=(\alph*)]
\item Conditional on $C=c$, the release $Y\sim M_c(\cdot\mid X)$ has the
advertised guarantee.
\item If $C$ is independent of $X$, the joint release $(C,Y)$ is
$(\varepsilon,\delta)$-LDP on $\mathcal R$; consequently its marginal mixture
uses the fixed weights $P_C(c)$ and has the same guarantee.
\item If $C$ depends on $X$, part (b) need not hold. If the release of $C$ is
itself $(\varepsilon_C,\delta_C)$-LDP, sequential composition gives
$(\varepsilon_C+\varepsilon,\delta_C+\delta)$ for $(C,Y)$.
\item At round $r$, selecting a channel measurably from prior privatized
history preserves the conditional per-round guarantee for a fresh user's
$X_r$, provided that user's record did not affect the history. Repeated
participation by the same user requires composition across that user's
releases.
\end{enumerate}
If the available relations depend on context, the same joint statement applies
only to a common relation contained in every $\mathcal R_c$.
\end{proposition}
\begin{proof}
Part (a) is the assumed guarantee of $M_c$. Under independence, for an event
$A$ in the joint output space, write $A_c=\{y:(c,y)\in A\}$. Then, for
$(x,x')\in\mathcal R$,
\begin{align}
\Pr((C,Y)\in A\mid X=x)
=\sum_c P_C(c)M_c(A_c\mid x)
\le e^\varepsilon\sum_cP_C(c)M_c(A_c\mid x')+\delta,
\end{align}
which proves (b). Dependence of $C$ on $X$ changes the mixing weights and can
itself reveal $X$; the stated bound then follows from standard adaptive
sequential composition. Conditional on history generated without the fresh
record, part (d) repeats the fixed-context argument. Reusing one record invokes
the same composition theorem.
\end{proof}

Thus a deployment can choose $t=t(C)$ and $(p,q)=(p(C),q(C))$ from the feasible set of Theorem~\ref{thm:joint-aistats} when $C$ is genuinely public and independent of the current protected value. This proposition does not justify tuning directly on an unprotected private symbol.
\subsection{Risk under distribution shift}
The exact risk formula also gives an immediate way to evaluate distribution shift. If the mechanism is designed under a nominal distribution $\pi_0$ but deployed under $\pi_1$, the risk is not an approximation: it is exactly \eqref{eq:exact-risk} with $\pi=\pi_1$ and $m=\pi_1P$. Consequently, shift evaluation requires no new privacy analysis; the privacy constraint is channel-level, while the risk changes with the input law.
This separation is useful for robust learning. Privacy feasibility is controlled by $(\varepsilon,\delta,H)$ and the channel parameters, whereas estimation risk is controlled by the interaction of the channel spectrum and the deployment distribution. The experiments below exploit this separation to compare nominal, risk-aware, and robust calibrations under changing distributions.

\section{Numerical Experiments}

The experiments are exact evaluations. They are not empirical privacy tests and they do not use Monte Carlo to estimate privacy or risk. For line supports we evaluate Corollary~\ref{cor:range-defect}; for finite cycles and nonhomogeneous paths we evaluate the corresponding finite hockey-stick sums directly and cross-check the cycle against Proposition~\ref{prop:cyclic-defect}. Risk is computed from Theorem~\ref{thm:linear-risk} or the algebraically equivalent quadratic identity in Theorem~\ref{thm:projection-risk}.

\paragraph{Evaluation protocol}
All privacy quantities are range restricted. For the translation-invariant
line and cycle channels, we report
$\delta_H^M=\max_{1\le h\le H}\Delta_\varepsilon^M(0,h),$
with $h$ interpreted modulo the cycle when applicable. For the
nonhomogeneous path, translation invariance is absent, so we instead maximize
over every ordered protected pair exactly as in \eqref{eq:path-defect}. These
are $(\varepsilon,\delta_H^M)$ guarantees on the stated relation
$\mathcal R_H$, not global all-pairs guarantees on the integer line. This
distinction is essential: every finite-radius line mechanism has global defect
one if arbitrary distant inputs are compared.

Unless otherwise stated, we set $(\varepsilon,H,\delta)=(1,3,0.45)$. The relatively large value $\delta=0.45$ is illustrative: it makes competing support and spectral designs visible, but it should not be read as a recommendation for a stringent privacy deployment. Experiment~1(b) therefore also sweeps $\delta$ down to $0.02$. For distribution estimation, we use a cyclic alphabet of size $K=31$ and a fixed nonuniform distribution with three localized modes. The cyclic model avoids boundary corrections and makes the channel circulant. Privacy is nevertheless evaluated on the finite cycle itself, because the line formula ceases to be exact when a support nearly fills the alphabet.

\paragraph{Reproducibility specification}
For $i\in\{0,\ldots,K-1\}$, define the normalized discrete bump
\[
g_{K,\mu,\sigma}(i)=
\frac{\exp\{-\tfrac12((i-\mu)/\sigma)^2\}}
{\sum_{j=0}^{K-1}\exp\{-\tfrac12((j-\mu)/\sigma)^2\}}.
\]
The $K=31$ design distribution is
$0.45g_{31,7,1.4}+0.35g_{31,15,1.9}+0.20g_{31,24,1.2}$.
Every cyclic search enumerates $t=1,\ldots,15$. For each radius, a uniform
grid of 3001 values of $q$ on $[0,1/(2t+1)]$ locates every privacy-feasible
component; its boundary is refined with Brent's method to absolute and
relative tolerances $10^{-14}$. Each feasible interval is split at every
analytically computed Fourier-eigenvalue zero, and bounded scalar risk
minimization is run on every nonsingular segment with absolute tolerance
$10^{-12}$ and at most 1000 iterations. Candidates with
$\min_\ell|\lambda_\ell|<10^{-10}$ are treated as singular, and exact privacy
comparisons use tolerance $10^{-11}$. 

\begin{figure}[t]
\centering
\includegraphics[width=0.90\linewidth]{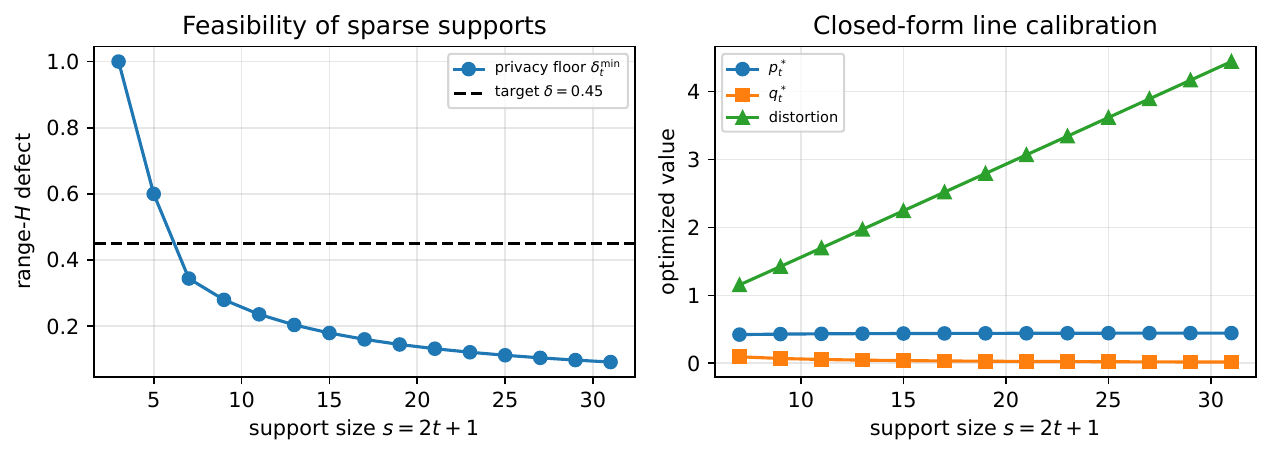}
\caption{Joint sparse randomized-response design for $(\varepsilon,H,\delta)=(1,3,0.45)$. Panel (a) distinguishes feasible and infeasible support sizes using the exact fixed-radius privacy floor. Panel (b) shows the closed-form probabilities and reporting distortion for feasible radii.}
\label{fig:aistats-joint-design}
\end{figure}

\paragraph{Experiment 1: privacy feasibility and optimized parameters}
Figure~\ref{fig:aistats-joint-design}(a) shows the minimum possible range-$H$ defect at each support size after optimizing over $p,q$. The dashed line is the target $\delta$. Supports below the line are feasible; supports above it cannot meet the target for any truthful probabilities. Figure~\ref{fig:aistats-joint-design}(b) shows the optimized $p_t^*(\delta)$, $q_t^*(\delta)$, and reporting distortion $R_1^{\RR}$ for feasible radii. The selected radius is the one minimizing $q_t^*(\delta)t(t+1)$, not necessarily the smallest feasible radius.

\begin{figure}[t]
\centering
\includegraphics[width=0.5\linewidth]{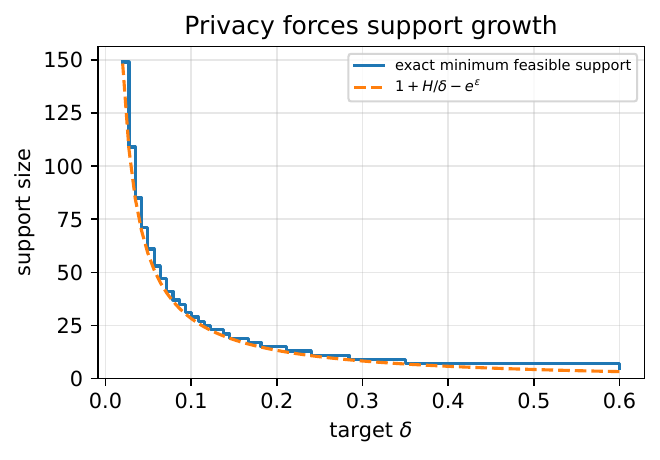}
\caption{Exact minimum feasible support size and the analytical lower bound
$1+H/\delta-e^\varepsilon$ for $(\varepsilon,H)=(1,3)$.}
\label{fig:support-scaling}
\vspace{-4mm}
\end{figure}

\paragraph{Experiment 1(b): privacy-imposed support scaling}
Figure~\ref{fig:support-scaling} varies the target $\delta$ while fixing
$(\varepsilon,H)=(1,3)$ on 80 equally spaced values in $[0.02,0.60]$, with
radii searched over $t=1,\ldots,499$. The exact minimum feasible support has the expected
integer staircase, while the analytical converse in
Corollary~\ref{cor:support-lower-bound} tracks its $H/\delta$ growth in the
strong-privacy regime. This sweep places the illustrative default
$\delta=0.45$ in context.

\begin{figure}[t]
\centering
\includegraphics[width=0.90\linewidth]{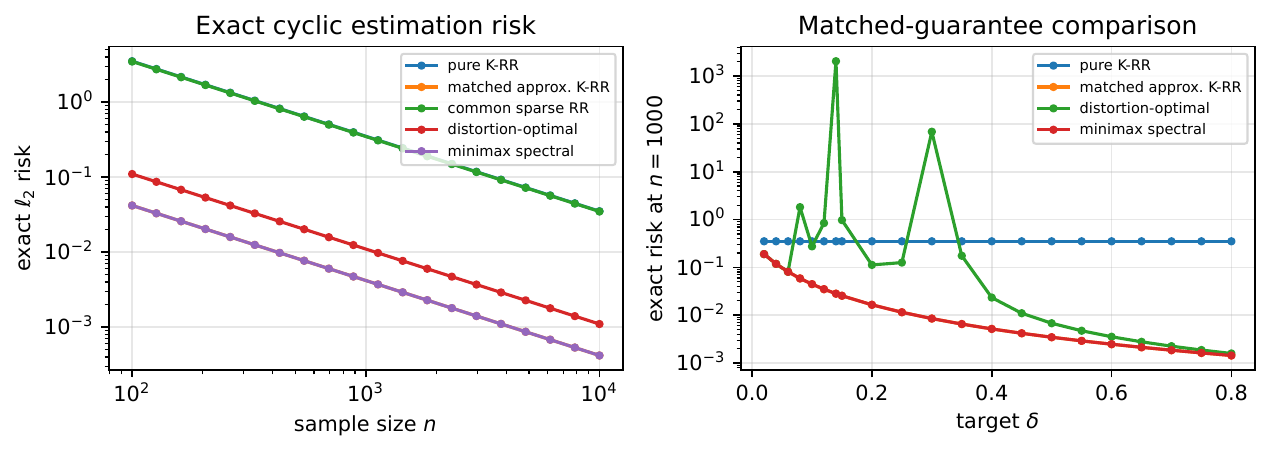}
\caption{Exact distribution-estimation risk on a cyclic alphabet with $K=31$. Panel (a) plots risk versus sample size for five mechanisms. Panel (b) fixes $n=1000$ and varies $\delta$ using matched privacy guarantees.}
\label{fig:aistats-risk}
\end{figure}

\paragraph{Experiment 2: finite-sample estimation risk}
We compare five mechanisms: pure full-support $K$-ary randomized response; a full-support randomized-response channel matched to the same $(\varepsilon,\delta,H)$ guarantee as the sparse mechanisms; common-calibrated sparse randomized response with $p=e^\varepsilon q$; the reporting-distortion optimum from Theorem~\ref{thm:joint-aistats}; and the cyclic minimax spectral design from Theorem~\ref{thm:cyclic-minimax}. The matched approximate baseline is essential: comparing an approximate relational mechanism only with pure all-pairs $K$-RR would confound mechanism geometry with a strictly stronger privacy requirement. For each candidate we form the channel matrix, verify invertibility using Proposition~\ref{prop:cyclic-spectrum}, evaluate privacy on the finite cycle, and compute the exact risk. Figure~\ref{fig:aistats-risk}(a) shows the resulting $1/n$ decay at the distinct integer values obtained by rounding down 20 logarithmically spaced points from $10^2$ to $10^4$.

\paragraph{Experiment 3: varying the approximate privacy budget}
Figure~\ref{fig:aistats-risk}(b) fixes $n=1000$ and evaluates the target
$\delta$ at seven equally spaced values from $0.02$ to $0.14$ and fourteen
equally spaced values from $0.15$ to $0.80$. Pure $K$-RR does not change with $\delta$, whereas the matched full-support channel, the distortion-optimal sparse design, and the minimax spectral design all use the same approximate relational privacy target. The distortion-optimal channel can exhibit nonmonotone estimation risk when a support transition places a Fourier eigenvalue near zero. The spectral design directly minimizes the exact minimax risk and avoids those instabilities. On the cycle its nominal and minimax versions coincide, as required by Theorem~\ref{thm:cyclic-minimax}.

\begin{figure}[t]
\centering
\includegraphics[width=0.82\linewidth]{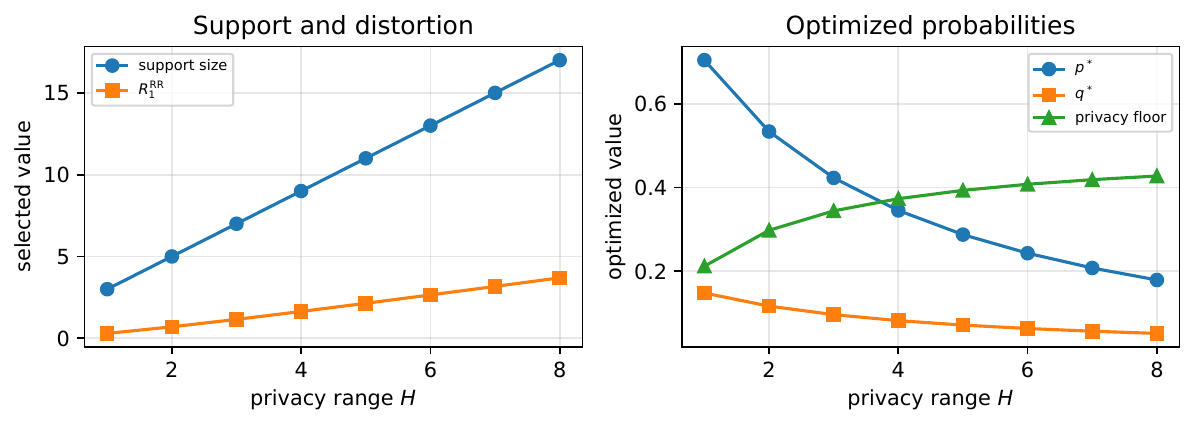}
\caption{Sensitivity of the optimized sparse randomized-response design to the privacy range $H$ at $(\varepsilon,\delta)=(1,0.45)$. Larger privacy ranges require larger supports and higher reporting distortion.}
\label{fig:aistats-range}
\end{figure}

\paragraph{Experiment 4: sensitivity to the privacy range}
Finally, Figure~\ref{fig:aistats-range} varies the integer privacy range over
$H=1,\ldots,8$ while holding $(\varepsilon,\delta)=(1,0.45)$ fixed and searches
$t=1,\ldots,79$. Larger $H$ requires more overlap between supports, so the optimized support size increases. The distortion also increases because more off-truth probability must be allocated over a larger neighborhood to satisfy the range-$H$ constraint. This illustrates that $H$ is a statistical modeling choice: choosing it too large moves the mechanism toward full-support privacy, while choosing it too small gives a weaker privacy relation.

\begin{table}[t]
\centering
\caption{Exact mechanism comparison for $K=31$, $(\varepsilon,H,\delta)=(1,3,0.45)$, and $n=1000$.}
\label{tab:selected-mechanisms}
\small
\begin{tabular}{lrrrrrr}
\toprule
Mechanism & $s$ & $p$ & $q$ & Defect & Dist. & Risk \\
\midrule
Pure full $K$-RR & 31 & 0.0831 & 0.03056 & $<10^{-15}$ & 7.3353 & 0.350838 \\
Matched full approx. RR & 31 & 0.4957 & 0.01681 & 0.4500 & 4.0344 & 0.004184 \\
Common sparse RR & 7 & 0.3118 & 0.11470 & 0.3441 & 1.3764 & 0.347988 \\
Distortion-opt. sparse RR & 7 & 0.4229 & 0.09618 & 0.4500 & 1.1542 & 0.010988 \\
Minimax spectral RR & 31 & 0.4957 & 0.01681 & 0.4500 & 4.0344 & 0.004184 \\
\bottomrule
\end{tabular}
\vspace{-6mm}
\end{table}

Table~\ref{tab:selected-mechanisms} summarizes the selected mechanisms and reports the exact attained defect rather than only the target. The common-calibrated and distortion-optimal sparse mechanisms have the same support size, but the former is poorly conditioned for inversion. The distortion-optimal design lowers $q$, increases the truthful mass, and improves risk substantially. The minimax spectral design sacrifices reporting-distance optimality by using full support. It coincides, to numerical precision, with the matched full-support approximate channel and achieves the smallest inverse-estimation risk. This equality is a consequence of the exact cyclic optimization, not evidence that a nominal prior changed the robust solution.

\begin{figure}[t]
\centering
\includegraphics[width=0.90\linewidth]{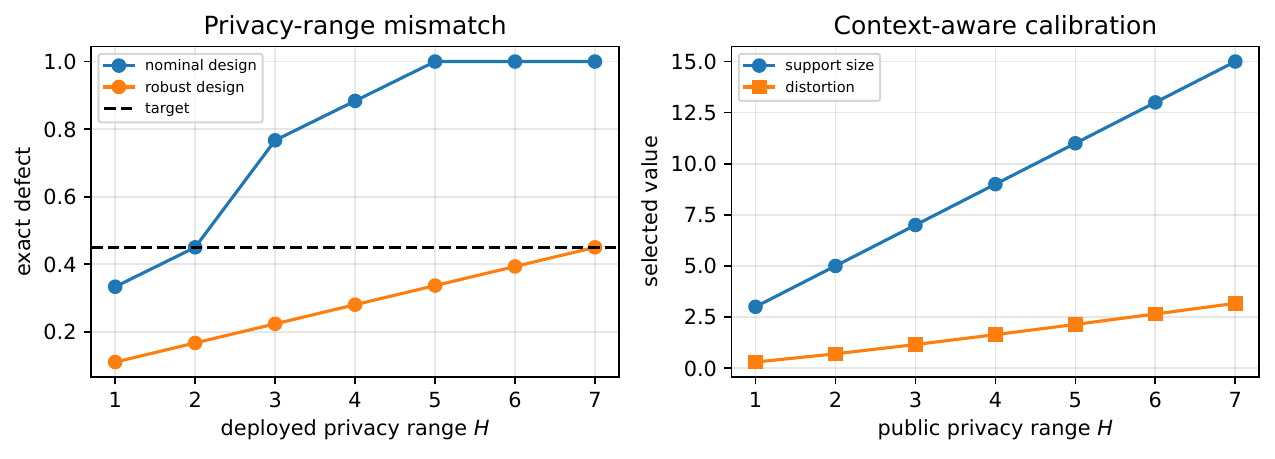}
\caption{Robustness and safe public-context adaptation. Left: exact defect under protected-range mismatch. Right: support size and distortion selected conditionally on an independent public range.}
\label{fig:jsait-robustrange}
\end{figure}

\paragraph{Experiment 5: robustness to privacy-range mismatch and public-context adaptation}
The previous experiments fix the privacy range $H$. In robust deployments, the range may be uncertain or context-dependent. Figure~\ref{fig:jsait-robustrange}(a) compares a mechanism calibrated at $H_0=2$ with one calibrated at $H_{\max}=7$. The nominal design reaches defect $0.45$ at $H=2$ but exceeds the target for every larger protected range; the $H_{\max}$ design remains below the target for all $H\le7$. Figure~\ref{fig:jsait-robustrange}(b) illustrates conditional calibration when the range is genuine public context. Proposition~\ref{prop:public-context-adaptation} applies only under its stated independence or composition conditions.

\begin{figure}[t]
\centering
\includegraphics[width=0.90\linewidth]{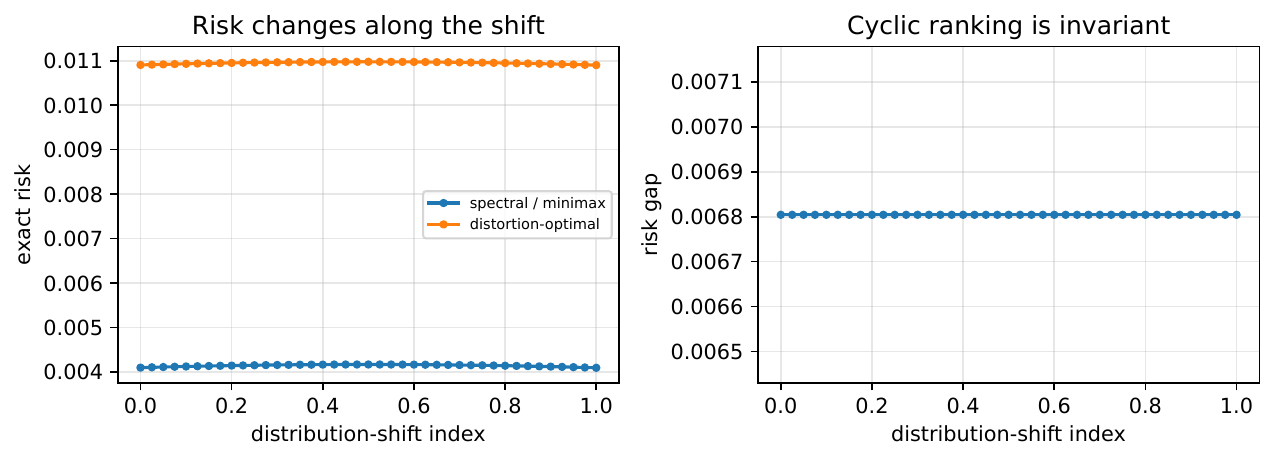}
\caption{Cyclic distribution-shift diagnostic. Left: exact risks change along the shift. Right: the inter-mechanism risk gap is invariant, so nominal and simplex-minimax calibration select the same cyclic channel.}
\label{fig:jsait-shift}
\end{figure}

\paragraph{Experiment 6: estimation risk under distribution shift}
Define
$\pi_L=0.70g_{31,4,1.0}+0.30g_{31,12,1.8}$ and
$\pi_R=0.25g_{31,10,1.5}+0.75g_{31,26,1.1}$. Figure~\ref{fig:jsait-shift}(a)
evaluates exact risk for
$\pi_\alpha=(1-\alpha)\pi_L+\alpha\pi_R$ at 41 equally spaced values of
$\alpha\in[0,1]$. Privacy is unchanged because it is a channel property, while each risk changes through the common term $-\|\pi\|_2^2/n$. Panel (b) shows that the risk gap between the distortion-optimal and spectral channels is constant along the path. Optimizing at either endpoint and optimizing over the full simplex return the same spectral mechanism. This corrects the earlier interpretation of the cyclic experiment: it illustrates the invariance theorem, not a nominal-versus-robust mechanism change.

\begin{figure}[t]
\centering
\includegraphics[width=0.90\linewidth]{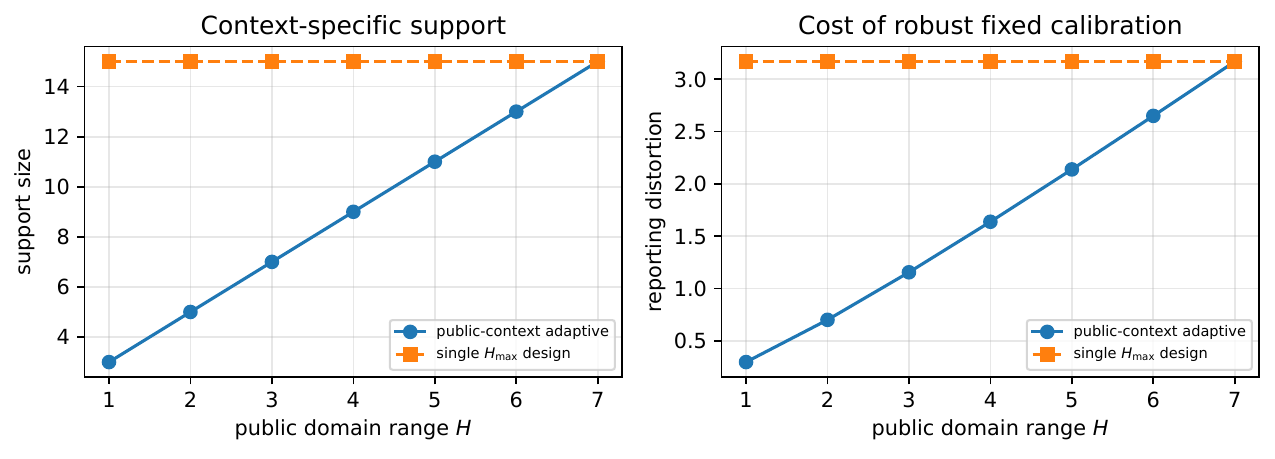}
\caption{Selection across independent public domains. Context-specific calibration uses smaller supports and lower distortion when the public protected range is small; the fixed $H_{\max}$ design is conservative in every domain.}
\label{fig:jsait-adaptivedomains}
\end{figure}

\paragraph{Experiment 7: adapting across public domains}
Figure~\ref{fig:jsait-adaptivedomains} gives a domain-level comparison. Each public domain has a different protected range and hence a different feasible support-radius set. Conditional calibration chooses the least-distorting feasible radius for that public range, whereas a single robust channel calibrated at $H_{\max}=7$ uses support size $15$ and distortion $3.1693$ in every domain. The comparison quantifies the utility cost of one-size-fits-all range robustness. It is not data-dependent tuning on an individual's current private value.

\paragraph{Experiment 8: a strict robust tradeoff on a nonhomogeneous path}
For the path channel in \eqref{eq:path-channel}, we take $K=11$,
$(\varepsilon,H,\delta)=(1,2,0.45)$, and $n=1000$. The disclosed candidate
set uses $t\in\{1,\ldots,5\}$, $\eta\in\{0,0.1,\ldots,1\}$, 45
logarithmically spaced values of $b\in[1.05,20]$, and
$\beta\in\{-0.4,-0.2,0,0.2,0.4\}$. Every candidate is screened by the exact
defect \eqref{eq:path-defect} with feasibility tolerance $10^{-11}$; matrices
with nonfinite condition number or $\operatorname{cond}_2(P)>10^8$ are
excluded before risk evaluation.

\begin{figure}[t]
\centering
\includegraphics[width=0.90\linewidth]{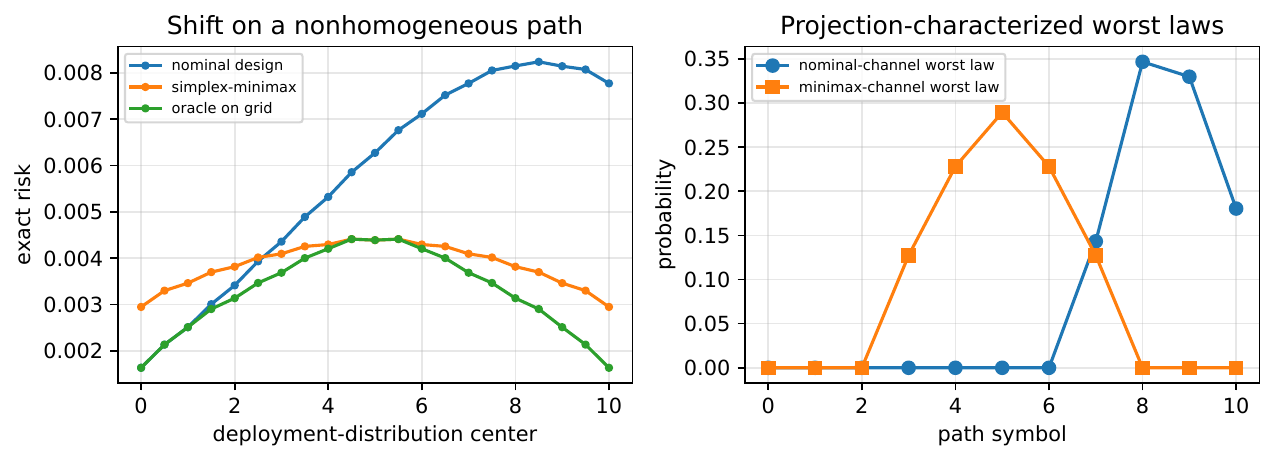}
\caption{Nonhomogeneous path experiment. Left: exact risk under distribution
shift, including a distribution-specific oracle over the disclosed candidate
family. Right: least-favorable laws from Theorem~\ref{thm:projection-risk}.}
\label{fig:jsait-nonhomogeneous}
\vspace{-4mm}
\end{figure}

The nominal law is $g_{11,0,0.35}$ and is concentrated near the left boundary.
The shift plot uses $g_{11,c,0.60}$ at
$c\in\{0,0.5,\ldots,10\}$. At each center, the distribution-specific oracle
is the minimum exact risk over the same screened candidate grid; it is not an
unconstrained lower bound. The nominal optimizer
selects $(t,\eta,b,\beta)=(4,0.3,4.008,-0.4)$, with exact defect $0.4428$,
nominal risk $1.180\times10^{-3}$, and simplex worst-case risk
$8.311\times10^{-3}$. The simplex-minimax optimizer selects
$(5,0.1,6.405,0)$, with exact defect $0.4438$, nominal risk
$2.575\times10^{-3}$, and worst-case risk $4.570\times10^{-3}$. Thus robust
calibration lowers worst-case risk by approximately $45\%$ under essentially
matched privacy, at the expected cost of larger nominal boundary risk.
Figure~\ref{fig:jsait-nonhomogeneous} shows the crossing risk curves and the
projection-characterized least-favorable laws. This is the example in which
robust calibration genuinely changes the mechanism.

\section{Discussion: Other support-constrained mechanisms}
The support-mismatch identity in \eqref{eq:decomp-aistats} is channel-agnostic,
so the same accounting principle applies beyond randomized response. For a
generic kernel
$
M_{\theta,t}(y\mid x)
\propto K_\theta(x,y)\1\{y\in S_t(x)\},
$
the support radius controls the mismatch region, whereas the kernel parameter
$\theta$ controls concentration and likelihood-ratio excess on the overlap.
The exact privacy defect can therefore be evaluated by summing the mismatch
mass and the positive overlap excess, after which the radius and kernel
parameters can be calibrated jointly.
For a radius-truncated Laplace kernel on the translation-invariant line, the
inverse scale $\lambda$ concentrates mass near the private symbol. Because the
truncation normalizer is then independent of the input, the sufficient
condition $\lambda H\le\varepsilon$ eliminates overlap excess throughout the
protected range, leaving boundary-tail mass to be charged to $\delta$. On a
finite path, row-normalizer ratios must also enter the overlap loss. Detailed Laplace and Gaussian
privacy--sparsity analyses appear in \citep{zamani2026sparseLDP}. A truncated
Gaussian kernel has a different geometry because its overlap privacy loss is
quadratic: increasing the support reduces boundary leakage but can require a
larger variance to control the likelihood ratio. Thus support and scale are
more tightly coupled than in the Laplace family.
Staircase channels replace a parametric decay law by optimizable shell
probabilities. At a fixed radius, normalization, monotonicity, reporting cost,
and the positive-part privacy constraints can be represented by a finite
linear program, connecting the present framework to extremal and staircase
mechanism design \citep{geng2014optimal,kairouz2014extremal}. Score-based and
exponential channels instead use a concentration parameter multiplying a task
score \citep{mcsherry2007mechanism}; their privacy and reporting cost remain
finite-sum quantities at each radius. For any such finite square channel that
is invertible, the exact inverse-risk and projection characterizations in
Theorems~\ref{thm:linear-risk}--\ref{thm:projection-risk} continue to apply.
What changes across mechanism families is the privacy-feasible channel set,
not the downstream risk identity. These observations identify natural
extensions, but the closed-form privacy--distortion optimization and numerical
claims of this paper concern sparse randomized response.
\section{Conclusion}

Support constraints turn local randomization into a joint geometric and
spectral design problem. Exact support leakage determines relational privacy
feasibility, while inverse-channel geometry determines finite-sample and
worst-case estimation risk. The projection theorem handles arbitrary closed
convex distribution uncertainty; cyclic symmetry yields an exact minimax
theorem, spectral converse, and a diagnostic against spurious prior-dependent
robustness claims. The nonhomogeneous path construction then shows that robust
calibration can materially change the mechanism once symmetry is removed. The
result is a precise, reproducible framework for robust support-constrained
private discrete learning, together with explicit conditions under which
context-dependent channel selection is safe.

\appendices

\section{Proofs of the Main Results}\label{app:missing-proofs}
This appendix provides the proofs of the results stated in the main paper. 

\subsection{Proof of Proposition~\ref{prop:pure-obstruction}}
Fix a pair $(x,x')$ in the privacy relation $\mathcal R$.  Suppose that there exists an output $y\in S(x)\setminus S(x')$.  By definition of the support, $M(y\mid x)>0$ and $M(y\mid x')=0$.  Applying pure $\varepsilon$-LDP to the event $A=\{y\}$ would require
$
        M(y\mid x)\le e^\varepsilon M(y\mid x')=0,
$
which is impossible.  Hence $S(x)\subseteq S(x')$ for every ordered pair to which the pure privacy constraint is applied.  If the relation is symmetric, applying the same argument to $(x',x)$ gives $S(x')\subseteq S(x)$, and therefore $S(x)=S(x')$.
For a finite-radius line mechanism with supports $S_t(x)=\{x-t,\ldots,x+t\}$ on an unbounded integer alphabet, choose two inputs with $|x-x'|>2t$.  Then $S_t(x)\cap S_t(x')=\varnothing$.  In particular, every $y\in S_t(x)$ lies in $S_t(x)\setminus S_t(x')$, so the argument above rules out finite pure all-pairs LDP.
\subsection{Proof of Proposition~\ref{prop:rr-defect-aistats}}
By translation invariance, it is enough to compare input $0$ with input $h\ge1$.  Write
\[
        S_0:=S_t(0)=\{-t,\ldots,t\},
        \qquad
        S_h:=S_t(h)=\{h-t,\ldots,h+t\}.
\]
The defect is computed from the decomposition
\[
\Delta_\varepsilon^{\RR}(0,h)
= M^{\RR}(S_0\setminus S_h\mid 0)
+
\sum_{y\in S_0\cap S_h}
\pos{M^{\RR}(y\mid0)-e^\varepsilon M^{\RR}(y\mid h)}.
\]

If $h>2t$, then $S_0$ and $S_h$ are disjoint.  Taking the event $S_0$ gives probability one under input $0$ and probability zero under input $h$, so the ordered defect is $1$.
Assume next that $1\le h\le t$.  Then,
$
        S_0\setminus S_h=\{-t,-t+1,\ldots,h-t-1\},
$
which has cardinality $h$.  Since $h-t-1\le -1$, this support-mismatch region does not contain the truthful symbol $0$.  Each point in it therefore has probability $q$ under input $0$, so the support-leakage contribution is $hq$.  On the overlap, the only point that can have larger probability under input $0$ than under input $h$ by more than the factor $e^\varepsilon$ is the truthful symbol $y=0$: it has probabilities $p$ under input $0$ and $q$ under input $h$.  Its contribution is $\pos{p-e^\varepsilon q}$.  At $y=h$, the contribution is $\pos{q-e^\varepsilon p}=0$ because $p\ge q$ and $e^\varepsilon\ge1$.  At every other overlap point both conditional probabilities are $q$, so the contribution is $\pos{q-e^\varepsilon q}=0$.  This gives
$
        \Delta_\varepsilon^{\RR}(0,h)=hq+\pos{p-e^\varepsilon q}.
$

Finally, suppose $t<h\le2t$.  The support-mismatch region is again $\{-t,\ldots,h-t-1\}$ and has cardinality $h$, but now it contains the truthful output $0$.  Thus its contribution is $p$ for $y=0$ plus $q$ for the remaining $h-1$ points, namely $p+q(h-1)$.  In this regime the overlap contains neither truthful symbol of the two inputs; all common outputs are nontruthful for both inputs and hence have probabilities $q$ and $q$.  Their positive-part contributions are zero. Therefore,
$
        \Delta_\varepsilon^{\RR}(0,h)=p+q(h-1),
$
which completes the three cases.
\subsection{Proof of Corollary~\ref{cor:range-defect}}
The range-$H$ defect is $\max_{1\le h\le H}\Delta_\varepsilon^{\RR}(0,h)$.  If $H>2t$, then the range contains a separation $h>2t$, and Proposition~\ref{prop:rr-defect-aistats} gives defect $1$ for that separation.  Since an ordered hockey-stick defect is always at most one, the range defect is $1$.
If $H\le t$, all allowed separations fall in the second case of Proposition~\ref{prop:rr-defect-aistats}.  The term $\pos{p-e^\varepsilon q}$ is independent of $h$, while $hq$ is nondecreasing in $h$.  Hence the maximum is attained at $h=H$, giving
$
        \delta_t(p,q;\varepsilon,H)=Hq+\pos{p-e^\varepsilon q}.
$
It remains to consider $t<H\le2t$.  For $h\le t$, the defect is $hq+\pos{p-e^\varepsilon q}$; for $t<h\le H$, it is $p+q(h-1)$, which is nondecreasing in $h$.  Thus the largest value in the second regime is at $h=H$.  This value also dominates the value at $h=t$: if $p-e^\varepsilon q\le0$, then $p+q(H-1)\ge tq$; if $p-e^\varepsilon q>0$, then
$
        p+q(H-1)-\bigl(tq+p-e^\varepsilon q\bigr)
        =q(H-1-t+e^\varepsilon)\ge0,
$
using $H\ge t+1$ and $e^\varepsilon\ge1$.  Hence the range defect is
$
        p+q(H-1)=1-2tq+q(H-1)=1-(2t-H+1)q.
$
\subsection{Proof of Theorem~\ref{thm:joint-aistats}}
For a fixed radius $t$, the normalization and truthful-bias constraints are
$
        p=1-2tq,
        \
        0\le q\le \frac{1}{2t+1}.
$
The reporting-distance objective is $R_1^{\RR}(p,q,t)=qt(t+1)$, so for fixed $t$ it is strictly increasing in $q$.  Therefore, once the privacy constraint is converted into a feasible interval for $q$, the distortion-optimal choice is the smallest feasible $q$.

First suppose $H\le t$.  Substituting $p=1-2tq$ into Corollary~\ref{cor:range-defect} gives
\[
\delta_t(q)=Hq+\pos{1-(2t+e^\varepsilon)q}.
\]
Let $a:=2t+e^\varepsilon$.  For $0\le q\le 1/a$, the positive part is active and
$
        \delta_t(q)=1-(a-H)q,
$
which is decreasing in $q$.  For $q\ge 1/a$, the positive part is zero and
$
        \delta_t(q)=Hq,
$
which is increasing in $q$.  Since $e^\varepsilon\ge1$, we have $1/a\le 1/(2t+1)$, so the kink lies in the truthful-bias domain.  The minimum fixed-radius defect is therefore attained at $q=1/a$ and equals
$
        \delta_t^{\min}=\frac{H}{2t+e^\varepsilon}.
$
Thus the radius is feasible exactly when $\delta\ge H/(2t+e^\varepsilon)$.  On the decreasing branch, the smallest $q$ that reaches the target $\delta$ solves
$
        1-(2t+e^\varepsilon-H)q=\delta,
$
so
$
        q_t^*(\delta)=\frac{1-\delta}{2t+e^\varepsilon-H}.
$
This is the first case of \eqref{eq:qstar-aistats}.
Now suppose $t<H\le2t$.  Corollary~\ref{cor:range-defect} gives
$
        \delta_t(q)=1-(2t-H+1)q.
$
This expression is decreasing over the feasible domain $0\le q\le1/(2t+1)$.  Therefore the smallest possible defect is attained at $q=1/(2t+1)$ and equals
$
        1-\frac{2t-H+1}{2t+1}=\frac{H}{2t+1}.
$
The radius is feasible exactly when $\delta\ge H/(2t+1)$.  Solving
$
        1-(2t-H+1)q=\delta
$
for the smallest feasible $q$ gives
$
        q_t^*(\delta)=\frac{1-\delta}{2t-H+1},
$
which is the second case of \eqref{eq:qstar-aistats}.  If $H>2t$, the defect is identically $1$, so no target $\delta<1$ is feasible.
For every feasible radius, $p_t^*(\delta)=1-2tq_t^*(\delta)$.  The fixed-radius optimum has distortion $q_t^*(\delta)t(t+1)$. Feasible radii exist because, for $t\ge H$, the privacy floor $H/(e^\varepsilon+2t)$ tends to zero. Moreover, for every $t\ge H$,
\[
q_t^*(\delta)t(t+1)
=\frac{(1-\delta)t(t+1)}{2t+e^\varepsilon-H}
\longrightarrow\infty.
\]
Consequently, only finitely many radii can improve upon any fixed feasible candidate. The minimum over the countable set of feasible integer radii is therefore attained, which proves the one-dimensional search in \eqref{eq:global-search-aistats}.
\subsection{Proof of Corollary~\ref{cor:privacy-floors}}
The proof follows by minimizing the fixed-radius defect over the truthful-bias interval $0\le q\le1/(2t+1)$, exactly as in the proof of Theorem~\ref{thm:joint-aistats}.  If $H>2t$, the supports are disjoint for an allowed separation and the floor is $1$.  If $H\le t$, the defect is
$
        Hq+\pos{1-(2t+e^\varepsilon)q},
$
which is minimized at $q=1/(2t+e^\varepsilon)$, giving $H/(2t+e^\varepsilon)$.  If $t<H\le2t$, the defect is $1-(2t-H+1)q$, minimized at the largest truthful-bias value $q=1/(2t+1)$, giving $H/(2t+1)$.  These are exactly the three cases in \eqref{eq:privacy-floor}.  A target $\delta$ is feasible if and only if it is at least the corresponding minimum value.
\subsection{Proof of Corollary~\ref{cor:support-lower-bound}}
Suppose first that $t<H$. A radius with $H>2t$ has defect one and is
infeasible. In the remaining case $t<H\le2t$, Corollary~\ref{cor:privacy-floors}
gives
$
\delta_t^{\min}=\frac{H}{2t+1}
\ge \frac{H}{2H-1}>\delta.
$
Thus every feasible radius must satisfy $t\ge H$. In that regime feasibility
requires
$
\delta\ge\frac{H}{e^\varepsilon+2t}.
$
Rearranging and using $s=2t+1$ yields
$s\ge1+H/\delta-e^\varepsilon$.
\subsection{Proof of Proposition~\ref{prop:risk-aware}}
Fix a radius $t$.  The constraints $p+2tq=1$ and $p\ge q$ reduce the probability design to the one-dimensional interval
$
        0\le q\le \frac{1}{2t+1},
        \ p=1-2tq.
$
The exact range-$H$ privacy constraint is the scalar condition
$
        \delta_t(1-2tq,q;\varepsilon,H)\le\delta.
$
When $H\le t$, this condition is
$
        Hq+\pos{1-(2t+e^\varepsilon)q}\le\delta.
$
On the left branch, where the positive part is active, this is equivalent to
$
        q\ge \frac{1-\delta}{2t+e^\varepsilon-H}.
$
On the right branch, where the positive part is inactive, it is equivalent to $q\le \delta/H$.  Intersecting with the truthful-bias domain gives \eqref{eq:q-interval-case1}.  When $t<H\le2t$, the privacy constraint is
$
        1-(2t-H+1)q\le\delta,
$
which gives the lower bound in \eqref{eq:q-interval-case2}; the upper bound is again $1/(2t+1)$.  If the privacy floor exceeds $\delta$, the corresponding interval is empty.
For every fixed pair $(t,q)$ with an invertible channel matrix $P$, Theorem~\ref{thm:linear-risk} gives the exact risk in \eqref{eq:risk-aware-problem}. Hence the original constrained optimization over $(t,p,q)$ is equivalent to minimizing this exact risk over the finite union, indexed by $t\in\mathcal T$, of the feasible one-dimensional intervals $\mathcal Q_t(\delta)$, with $p$ determined by $p=1-2tq$. This proves the finite search formulation \eqref{eq:risk-aware-search}. On a finite cycle, Proposition~\ref{prop:cyclic-defect} supplies the exact scalar sublevel set in $q$; the same reduction then applies verbatim.
\subsection{Proof of Theorem~\ref{thm:linear-risk}}
We use row-vector notation.  The output distribution is $m=\pi P$.  Since $\hat m$ is the empirical distribution of $n$ independent outputs with law $m$, it satisfies
$
        \E[\hat m]=m,
        \
        \operatorname{Cov}(\hat m)=\frac{1}{n}\bigl(\diag(m)-m^Tm\bigr),
$
where $m^Tm$ denotes the outer product.  The estimator is $\hat\pi=\hat m P^{-1}$, so
$
        \E[\hat\pi]=\E[\hat m]P^{-1}=mP^{-1}=\pi PP^{-1}=\pi,
$
which proves unbiasedness.  Its covariance is obtained by the usual linear transformation rule:
\[
        \operatorname{Cov}(\hat\pi)
        =P^{-T}\operatorname{Cov}(\hat m)P^{-1}
        =\frac{1}{n}P^{-T}\bigl(\diag(m)-m^Tm\bigr)P^{-1}.
\]
For an unbiased vector estimator, the mean-squared $\ell_2$ error equals the trace of its covariance matrix.  Taking the trace gives \eqref{eq:exact-risk}.
\subsection{Proof of Theorem~\ref{thm:projection-risk}}
Cyclicity of trace and $m=\pi P$ give
\begin{align}
nR(\pi;P)
&=\tr\!\left(\diag(m)B\right)
-\tr\!\left(P^{-T}m^TmP^{-1}\right)\notag\\
&=md-\|mP^{-1}\|_2^2=\pi Pd-\|\pi\|_2^2
=\pi a(P)-\|\pi\|_2^2.
\end{align}
Completing the square yields
$
nR(\pi;P)=\frac{\|a(P)\|_2^2}{4}
-\left\|\pi^T-\frac{a(P)}{2}\right\|_2^2.
$
Maximizing over $\Pi$ is therefore equivalent to minimizing Euclidean
distance from $a(P)/2$. A nonempty closed subset of the finite-dimensional
simplex is compact, and projection onto a closed convex set is unique. This
proves \eqref{eq:projection-law} and \eqref{eq:projection-value}. The
Karush--Kuhn--Tucker conditions for projection onto $\Delta_K$ give
$\pi_i^*=\max\{a_i(P)/2-\tau,0\}$, with $\tau$ chosen so the coordinates sum
to one.
\subsection{Proof of Corollary~\ref{cor:risk-stability}}
From \eqref{eq:quadratic-risk},
\begin{align}
n|R(\pi;P)-R(\pi_0;P)|
&=\left|(\pi-\pi_0)a(P)
-(\pi+\pi_0)(\pi-\pi_0)^T\right|\notag\\
&\le\left(\|a(P)\|_\infty+\|\pi+\pi_0\|_\infty\right)
\|\pi-\pi_0\|_1\notag\\
&\le(\|a(P)\|_\infty+2)\|\pi-\pi_0\|_1.
\end{align}
For a total-variation ball of radius $\rho$,
$\|\pi-\pi_0\|_1\le2\rho$.
\subsection{Proof of Proposition~\ref{prop:cyclic-defect}}
Let $A=B_t(x)$ and $A'=B_t(x')$ be the two cyclic balls, each of cardinality
$s=2t+1$, and take $h=d_K(x,x')\le\lfloor K/2\rfloor$. Shifting a cyclic arc
of size $s$ by $h$ removes
$
|A\setminus A'|=\min\{h,s,K-s\}
=\min\{h,2t+1,K-2t-1\}
$
points: the loss grows as $h$ until it is capped either by the arc size or by
the complement size. If $h\le t$, the truthful output $x$ lies in the overlap,
so every mismatch point has mass $q$ and the only possible overlap excess is
$\pos{p-e^\varepsilon q}$. If $h>t$, the truthful output $x$ lies in the
mismatch. The term $qr_{K,t}(h)$ counts it as mass $q$, so the correction is
$p-q$, and no overlap point produces positive excess. Combining the two cases
gives \eqref{eq:cyclic-defect}. Maximizing over the protected cyclic distances
gives the range statement. When $2t+H<K$, neither the support nor complement
cap is active before the line-overlap cases are exhausted, recovering
Corollary~\ref{cor:range-defect}.
\subsection{Proof of Proposition~\ref{prop:cyclic-spectrum}}
On the cyclic alphabet $\mathbb Z_K$, the channel probability from $x$ to $y$ depends only on the offset $y-x$ modulo $K$.  Thus the channel matrix is circulant.  Its first row has value $p$ at offset $0$, value $q$ at offsets $\pm1,\ldots,\pm t$, and value $0$ at all other offsets.  Every circulant matrix is diagonalized by the discrete Fourier basis.  For frequency $\ell$, the eigenvalue is the Fourier transform of the first row:
$
        \lambda_\ell
        =p+q\sum_{j=1}^t e^{2\pi i\ell j/K}
          +q\sum_{j=1}^t e^{-2\pi i\ell j/K}.
$
Combining conjugate terms gives
$
        \lambda_\ell=p+2q\sum_{j=1}^t\cos\left(\frac{2\pi \ell j}{K}\right),
$
which is \eqref{eq:fourier-eigs}.  Since the Fourier matrix is invertible, the circulant channel is invertible if and only if none of its Fourier eigenvalues is zero.
\subsection{Proof of Theorem~\ref{thm:cyclic-minimax}}
For circulant $P$, the matrix $B=P^{-1}P^{-T}$ is circulant and every diagonal
entry equals
$
c(P)=\frac{1}{K}\tr(B)
=\frac{1}{K}\sum_{\ell=0}^{K-1}|\lambda_\ell|^{-2}.
$
Thus $d=c(P)\1$. Since a stochastic circulant matrix satisfies $P\1=\1$,
we have $a(P)=Pd=c(P)\1$. Substitution in \eqref{eq:quadratic-risk} gives
$
R(\pi;P)=\frac{1}{n}\left[c(P)-\|\pi\|_2^2\right].
$
Over the simplex, $\|\pi\|_2^2$ is uniquely minimized at the uniform law,
where it equals $1/K$. Because $\lambda_0=1$, substitution yields
\eqref{eq:uniform-risk}. Finally,
$
R(\pi;P_1)-R(\pi;P_2)=\frac{c(P_1)-c(P_2)}{n},
$
which is independent of $\pi$.
\subsection{Proof of Corollary~\ref{cor:spectral-sandwich}}
Cauchy--Schwarz gives
$
\sum_{\ell=1}^{K-1}|\lambda_\ell|^{-2}
\ge
\frac{(K-1)^2}{\sum_{\ell=1}^{K-1}|\lambda_\ell|^2}.
$
Parseval's identity applied to the first channel row gives
$
\sum_{\ell=0}^{K-1}|\lambda_\ell|^2=K(p^2+2tq^2).
$
Subtracting $|\lambda_0|^2=1$ and using \eqref{eq:uniform-risk} proves the
lower bound. The upper bound follows from
$|\lambda_\ell|^{-2}\le\mu(P)^{-2}$ for each of the $K-1$ nonzero
frequencies.
\bibliographystyle{IEEEtran}
\bibliography{JSAIT}

\begin{thebibliography}{99}
\bibitem[Acharya et~al.(2020)]{acharya2020context}
J. Acharya, K. Bonawitz, P. Kairouz, D. Ramage, and Z. Sun.
\newblock Context aware local differential privacy.
\newblock In \emph{Proc. 37th International Conference on Machine Learning}, pages 52--62. PMLR, 2020.

\bibitem[Acharya et~al.(2021)]{acharya2021sparse}
J. Acharya, P. Kairouz, Y. Liu, and Z. Sun.
\newblock Estimating sparse discrete distributions under privacy and communication constraints.
\newblock In \emph{Algorithmic Learning Theory}, pages 79--98. PMLR, 2021.

\bibitem[Agarwal et~al.(2021)]{agarwal2021skellam}
N. Agarwal, P. Kairouz, and Z. Liu.
\newblock The Skellam mechanism for differentially private federated learning.
\newblock In \emph{Advances in Neural Information Processing Systems}, volume 34, pages 5052--5064, 2021.

\bibitem[Asoodeh et~al.(2021)]{asoodeh2021ldpcontraction}
S. Asoodeh, M. Aliakbarpour, and F. P. Calmon.
\newblock Local differential privacy is equivalent to contraction of {$E_\gamma$}-divergence.
\newblock \emph{arXiv preprint arXiv:2102.01258}, 2021.

\bibitem[Asoodeh et~al.(2021)]{asoodeh2021threevariants}
S. Asoodeh, J. Liao, F. P. Calmon, O. Kosut, and L. Sankar.
\newblock Three variants of differential privacy: Lossless conversion and applications.
\newblock \emph{IEEE Journal on Selected Areas in Information Theory}, 2(1):208--222, 2021.



\bibitem[Chen et~al.(2022)]{chen2022poissonbinomial}
W.-N. Chen, A. {\"O}zg{\"u}r, and P. Kairouz.
\newblock The Poisson binomial mechanism for secure and private federated learning.
\newblock \emph{arXiv preprint arXiv:2207.09916}, 2022.

\bibitem[Chhor and Sentenac(2023)]{chhor2023robust}
J. Chhor and F. Sentenac.
\newblock Robust estimation of discrete distributions under local differential privacy.
\newblock In \emph{Proc. 34th International Conference on Algorithmic Learning Theory}, pages 411--446. PMLR, 2023.


\bibitem[Duchi et~al.(2014)]{duchi2014local}
J. C. Duchi, M. I. Jordan, and M. J. Wainwright.
\newblock Local privacy and minimax bounds: Sharp rates for probability estimation.
\newblock In \emph{Advances in Neural Information Processing Systems}, 2014.

\bibitem[Duchi et~al.(2018)]{duchi2018minimax}
J. C. Duchi, M. I. Jordan, and M. J. Wainwright.
\newblock Minimax optimal procedures for locally private estimation.
\newblock \emph{Journal of the American Statistical Association}, 113(521):182--201, 2018.

\bibitem[Dwork et~al.(2006)]{dwork2006calibrating}
C. Dwork, F. McSherry, K. Nissim, and A. Smith.
\newblock Calibrating noise to sensitivity in private data analysis.
\newblock In \emph{Theory of Cryptography Conference}, pages 265--284. Springer, 2006.

\bibitem[Dwork and Roth(2014)]{dwork2014book}
C. Dwork and A. Roth.
\newblock \emph{The Algorithmic Foundations of Differential Privacy}.
\newblock Foundations and Trends in Theoretical Computer Science, 9(3--4):211--487, 2014.

\bibitem[Erlingsson et~al.(2014)]{rappor2014}
U. Erlingsson, V. Pihur, and A. Korolova.
\newblock RAPPOR: Randomized aggregatable privacy-preserving ordinal response.
\newblock In \emph{Proc. ACM SIGSAC Conf. Computer and Communications Security}, pages 1054--1067, 2014.

\bibitem[Geng and Viswanath(2014)]{geng2014optimal}
Q. Geng and P. Viswanath.
\newblock The optimal mechanism in differential privacy.
\newblock In \emph{Proc. IEEE International Symposium on Information Theory}, pages 2371--2375, 2014.

\bibitem[Kairouz et~al.(2014)]{kairouz2014extremal}
P. Kairouz, S. Oh, and P. Viswanath.
\newblock Extremal mechanisms for local differential privacy.
\newblock In \emph{Advances in Neural Information Processing Systems}, 2014.

\bibitem[Kairouz et~al.(2016)]{kairouz2016discrete}
P. Kairouz, K. Bonawitz, and D. Ramage.
\newblock Discrete distribution estimation under local privacy.
\newblock In \emph{International Conference on Machine Learning}, pages 2436--2444. PMLR, 2016.

\bibitem[Joseph et~al.(2018)]{joseph2018evolving}
M. Joseph, A. Roth, J. Ullman, and B. Waggoner.
\newblock Local differential privacy for evolving data.
\newblock In \emph{Advances in Neural Information Processing Systems}, volume 31, 2018.

\bibitem[Li et~al.(2023)]{li2023robustness}
M. Li, T. B. Berrett, and Y. Yu.
\newblock On robustness and local differential privacy.
\newblock \emph{Annals of Statistics}, 51(2):717--737, 2023.

\bibitem[McSherry and Talwar(2007)]{mcsherry2007mechanism}
F. McSherry and K. Talwar.
\newblock Mechanism design via differential privacy.
\newblock In \emph{Proc. IEEE Symposium on Foundations of Computer Science}, pages 94--103, 2007.

\bibitem[Murakami et~al.(2019)]{murakami2019utility}
T. Murakami, and Y. Kawamoto.
\newblock Utility-optimized local differential privacy mechanisms for distribution estimation.
\newblock In \emph{USENIX Security Symposium}, 2019.

\bibitem[Pastore and Gastpar(2021)]{pastore2021randomized}
A. Pastore and M. Gastpar.
\newblock Locally differentially-private randomized response for discrete distribution learning.
\newblock \emph{Journal of Machine Learning Research}, 22(132):1--56, 2021.

\bibitem[Sadeghi et~al.(2020)]{sadeghi2020countqueries}
P. Sadeghi, S. Asoodeh, and F. d. P. Calmon.
\newblock Differentially private mechanisms for count queries.
\newblock \emph{arXiv preprint arXiv:2007.09374}, 2020.

\bibitem[Sadeghi and Chien(2024)]{sadeghi2024abs}
P. Sadeghi and C.-H. Chien.
\newblock On the connection between the {ABS} perturbation methodology and differential privacy.
\newblock \emph{Journal of Privacy and Confidentiality}, 14(2), 2024.

\bibitem[Sadeghi et~al.(2021)]{sadeghi2020privacywatchdog}
P. Sadeghi, N. Ding, and T. Rakotoarivelo.
\newblock On properties and optimization of information-theoretic privacy watchdog.
\newblock In \emph{Proc. IEEE Information Theory Workshop}, 2021.

\bibitem[Sima et~al.(2024)]{sima2024online}
J. Sima, C. Wu, O. Milenkovic, and W. Szpankowski.
\newblock Online distribution learning with local privacy constraints.
\newblock In \emph{Proc. 27th International Conference on Artificial Intelligence and Statistics}, pages 460--468. PMLR, 2024.

\bibitem[Warner(1965)]{warner1965randomized}
S. L. Warner.
\newblock Randomized response: A survey technique for eliminating evasive answer bias.
\newblock \emph{Journal of the American Statistical Association}, 60(309):63--69, 1965.

\bibitem[Zamani et~al.(2025)]{zamani2025infogeometric}
A. Zamani, P. Sadeghi, and M. Skoglund.
\newblock An information geometric approach to local information privacy with applications to max-lift and local differential privacy.
\newblock In \emph{Proc. IEEE Information Theory Workshop}, 2025.

\bibitem[Zamani et~al.(2026a)]{zamani2026sparsepointwise}
A. Zamani, S. Daei, P. Sadeghi, and M. Skoglund.
\newblock Sparse point-wise privacy leakage: Mechanism design and fundamental limits.
\newblock In \emph{2026 IEEE International Symposium on Information Theory (ISIT)}, Guangzhou, China, 2026, pp. 1--6.
\newblock \href{https://doi.org/10.1109/ISIT62367.2026.11653658}{doi:10.1109/ISIT62367.2026.11653658}.

\bibitem[Zamani et~al.(2026b)]{zamani2026sparseLDP}
A. Zamani, S. Daei, P. Sadeghi, and M. Skoglund.
\newblock Sparse discrete Laplace and Gaussian mechanisms under local differential privacy.
\newblock In \emph{Proc. IEEE Information Theory Workshop (ITW)}, 2026.

\end{thebibliography}
\end{document}